\documentclass[nonblindrev]{informs3_ejor}

\OneAndAHalfSpacedXI 

\usepackage{multibib}
\newcites{appendix}{References (supplements)}

\usepackage{natbib}
 \bibpunct[, ]{(}{)}{,}{a}{}{,}%
 \def\bibfont{\small}%
\usepackage{booktabs}
\usepackage[utf8]{inputenc}
\usepackage{lmodern}

\usepackage{siunitx}
\usepackage{amsmath}
\newcounter{subeq}
\renewcommand{\thesubeq}{\theequation\alph{subeq}}
\newcommand{\newsubeqblock}{\setcounter{subeq}{0}\refstepcounter{equation}}
\newcommand{\mysubeq}{\refstepcounter{subeq}\tag{\thesubeq}}

\usepackage{mathtools}
\DeclarePairedDelimiter{\ceil}{\lceil}{\rceil}
\DeclarePairedDelimiter{\floor}{\lfloor}{\rfloor}

\usepackage{subcaption}
\usepackage{float}
\usepackage{lscape}
\usepackage{mathtools,algorithm}
\usepackage[noend]{algpseudocode}
  \algtext*{EndFunction}
\usepackage{varwidth}
\usepackage[shortlabels]{enumitem}

\newcommand{\define}{\ensuremath\stackrel{\text{def}}{=}}
\newcommand{\abs}[1]{\left| #1 \right|}

\newcommand{\pns}{DNRP-LSDC\xspace}
\newcommand{\pnl}{distribution network reconfiguration problem with line-specific demand coincidence\xspace}

\usepackage{xspace}
  \newcommand\ie{i.\,e.,\xspace}
  
  \newcommand\eg{e.\,g.,\xspace}
  
	\newcommand\etc{etc.\xspace}
	\newcommand\cf{cf.\xspace}

\algnewcommand{\IIf}[1]{\State\algorithmicif\ #1\ \algorithmicthen}
\algnewcommand{\EndIIf}{\unskip\ \algorithmicend\ \algorithmicif}
\newlength{\maxwidth}
\newcommand{\algalign}[2]
{\makebox[\maxwidth][r]{$#1{}$}${}#2$}

\usepackage{xcolor}

\usepackage{todonotes}

\newcommand\MOVED[1]{{\color{black}#1}} 
\newcommand\REVTWO[1]{{\color{black}#1}} 
\newcommand\NEW[1]{{\color{black}#1}} 
\newcommand\TEMPERED[1]{{\color{black}#1}} 

\usepackage[normalem]{ulem}

\usepackage[capitalize,noabbrev,sort&compress]{cleveref}

\usepackage{graphicx}

\usepackage{listings}
\usepackage{colortbl}
\usepackage{color}

\usepackage{multirow}

\usepackage{csquotes}

\TheoremsNumberedThrough     
\ECRepeatTheorems

\EquationsNumberedThrough    

\MANUSCRIPTNO{} 


\def\FigureNoteName{Notes.}

\begin{document}


\RUNAUTHOR{Gust et al.}

\RUNTITLE{Designing Electricity Distribution Networks under Demand Coincidence}

\TITLE{Designing Electricity Distribution Networks:\\ The Impact of Demand Coincidence}


\ARTICLEAUTHORS{%
\AFF{Gunther Gust (corresponding author), Center for Artificial Intelligence and Data Science, University of Würzburg, Sanderring 2, 97070 Würzburg, Germany,  \EMAIL{gunther.gust@uni-wuerzburg.de}}
\AFF{Alexander Schlueter, University of Freiburg, Rempartstraße 10-16, 79098 Freiburg, Germany, \EMAIL{alexander.schlueter@is.uni-freiburg.de}}
\AFF{Stefan Feuerriegel, LMU Munich, Geschwister-Scholl-Platz 1, 80539 Munich, Germany,  \EMAIL{feuerriegel@lmu.de}}
\AFF{Ignacio Ubeda, Center for Artificial Intelligence and Data Science, University of Würzburg, Sanderring 2, 97070 Würzburg, Germany,  \EMAIL{ignacio.ubeda@uni-wuerzburg.de}}
\AFF{Jonathan T. Lee, University of California at Berkeley, Berkeley, CA	94720 USA, \EMAIL{jtlee@berkeley.edu}}
\AFF{Dirk Neumann, University of Freiburg, Rempartstraße 10-16, 79098 Freiburg, Germany, \EMAIL{dirk.neumann@is.uni-freiburg.de}}
} 

%
%

\ABSTRACT{%
\noindent
With the global effort to reduce carbon emissions, clean technologies such as electric vehicles and heat pumps are increasingly introduced into electricity distribution networks. These technologies considerably increase electricity flows and can lead to more coincident electricity demand. In this paper, we analyze how such increases in demand coincidence impact future distribution network investments. For this purpose, we develop a novel model for designing electricity distribution networks, called the \emph{\pnl}~(\pns). Our analysis is two-fold: (1)~We apply our model to a large sample of real-world networks from a Swiss distribution network operator. We find that a high demand coincidence due to, for example, a large-scale uptake of electric vehicles, requires a substantial amount of new network line construction and increases average network cost by 84\,\% in comparison to the status quo. (2)~We use a set of synthetic networks to isolate the effect of specific network characteristics. Here, we show that high coincidence has a more detrimental effect on large networks and on networks with low geographic consumer densities, as present in, \eg rural areas. \NEW{We also show that expansion measures are robust to variations in the cost parameters.} Our results demonstrate the necessity of designing policies and operational protocols that reduce demand coincidence. Moreover, our findings show that operators of distribution networks must consider the demand coincidence of new electricity uses and adapt investment budgets accordingly. Here, our solution algorithms for the \pns problem can support operators of distribution networks in strategic and operational network design tasks.
}%


\vspace{-2cm}
\maketitle


%

\sloppy
\raggedbottom

\vspace{-1cm}


\section{Introduction}


Increasing the sustainability of the electricity sector is an important lever in the global effort to reduce greenhouse gas emissions, which was recently manifested in the Paris Agreement \citep{UN.2015}. To achieve the sustainability goals, it is crucial that clean energy technologies---such as photovoltaic systems, heat pumps, and electric vehicles---receive further uptake. Designing future electricity distribution networks under these technologies, however, is challenging as they considerably impact electricity flows in distribution networks, both in terms of magnitude and temporal dynamics \citep{Parker.2019}. 

Due to changes in the temporal dynamics, it is expected that electricity distribution networks, which connect end-users to the higher layers of the electricity system, require substantial investment for capacity expansions  \citep[\eg][]{U.S.DepartmentofEnergy.2018}. For instance, charging electric vehicles can create new peaks of electricity flows, if such charging is conducted by several users in the network at the same point in time \citep{Gaul.2017,MCKINNEY2023,Salah.2015, Verzijlbergh.2011}. Similarly, smart grid technologies (such as local battery storages, heat pumps, and others) can shift demand to similar points in time, for example, when several consumers use these devices simultaneously---\eg to take advantage of low electricity prices \citep{Kahlen.2018, Strbac.2008}. More generally, the phenomenon of coincident electricity flows also occurs on the supply side. Here, photovoltaic systems generate electricity when solar irradiation is high (with peaks typically occurring during noon hours) and thus are also characterized by coincident flows.

\MOVED{Industry reports agree that electricity distribution networks require massive investments because of the transition to such decentralized sustainable technologies. For example, in the United States, annual investments in electricity distribution networks amounted to 25 billion dollars in 2017 and continue to increase \citep{U.S.DepartmentofEnergy.2018, U.S.DepartmentofEnergy.2020}.} \NEW{Similarly, European distribution networks are estimated to require around 400 billion euros of investments in the course of this transition until 2030 \citep{connectingthedots, Hoflich.2012}}. \NEW{Despite these large investment requirements, current network planning practice is still mostly reactive and based on subjective expert opinion \citep{Gust.2016, Gust.2017}.}\footnote{\NEW{Additional background information on the transition to more data-driven planning practices of utility companies is provided in, \eg \citet{Gust.2016, Gust.2017}, where the project that motivated this paper is described.}} Even less attention has been paid to the effect of the temporal dynamics within electricity demand.  

\begin{figure}[h!]
\begin{tabular}{cc}
\includegraphics[width=0.39\textwidth]{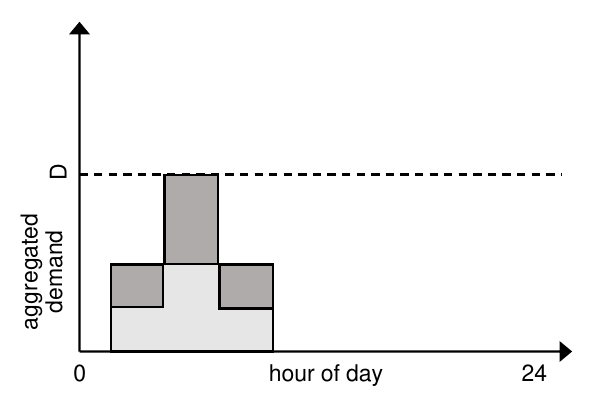} & \includegraphics[width=0.59\textwidth]{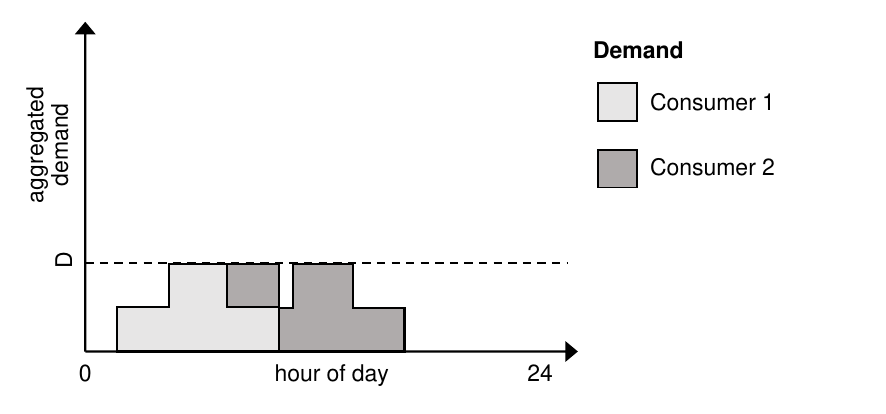} \\
\footnotesize High coincidence (here: $\gamma = 1$) & \footnotesize \hspace{-2cm}Low coincidence (here: $\gamma = 0.5$)
\end{tabular}
\vspace{0.2cm}
\caption{\raggedright Temporal electricity demand of two consumers with different degrees of demand coincidence. The aggregated peak demand $D$ in the electricity network depends on the coincidence factor $\gamma$. Left: Under a high demand coincidence, peak demands of consumers occur at similar points in time. Right: Under lower demand coincidence, the temporal overlap of the individual consumer peak demands is smaller.}
\label{fig:concidence_factor}
\end{figure}

Temporal overlaps in the electricity demand of consumers are commonly modeled using the so-called \emph{coincidence factor} \citep[cf.][]{Dickert.2010}. The coincidence factor $\gamma$  relates peak demand values of individual consumers to the overall peak demand of a group of consumers (as illustrated in \cref{fig:concidence_factor}). Hence, a larger coincidence factor implies that demand takes place at the same time. This, in turn, necessitates electricity networks that can serve large peak demands. Research, so far, has focused on estimating the coincidence factor of different technologies, such as electric vehicles, photovoltaic systems, heat pumps, and others \citep[\eg][]{Boait.2015, Konstantelos.2014, Verzijlbergh.2011}. For instance, electric vehicles can lead in the worst case to a fully coincident electricity demand, \ie $\gamma = 1$  \citep{Verzijlbergh.2011}. 

\MOVED{Such coincident demand has direct consequences for electricity networks. For example, it has been shown that more coincident demand generally requires larger line capacities \citep{Kaur.2008}. Coincident demand has previously also been linked to overloads on higher layers of the electricity systems---at the level of transmission networks \citep{Salah.2015} and network transformers \citep{Gwisdorf.2010}. The effect on the layer of electricity distribution networks has, however, to the best of our knowledge, not yet been quantified. In particular, the impact of demand coincidence on the investment costs for designing electricity distribution networks has been unclear.} 


We address this gap and thereby contribute to the literature in three ways:
\begin{enumerate}
\item To determine the impact of demand coincidence on network cost, we develop a novel model for designing electricity distribution networks: the \emph{\pnl}~(\pns). Our \pns extends the distribution network reconfiguration model \citep[\eg][]{Avella.2005} by including a variable coincidence factor. Furthermore, our \pns is suitable for greenfield network design, \ie it allows networks to be designed from scratch \REVTWO{and, unlike related brownfield planning approaches}, it is \emph{not} restricted to a limited amount of reconfiguration options. We provide exact solutions to the \pns for small network instances. For larger network instances, given the NP-hardness of the \pns, we develop solution heuristics. Our heuristics are designed in such a way that they leverage the unique physical properties of the \pns. 
We demonstrate the effectiveness of the heuristics by providing numerical upper and lower bounds for the exact solution based on simplified problem instances. We finally discuss how our methodological advances support operators of distribution networks in strategic and operational network design tasks.
\item We estimate the impact of demand coincidence on the investment cost of electricity distribution networks. We do so (a) in an evaluation based on a large sample of real-world networks and (b) in a numerical evaluation based on synthetic networks. We find that a larger coincidence factor---due to, \eg the diffusion of electric vehicles, heat pumps, and other technological innovations---increases the cost of real-world networks by, on average, 84\,\% and, in the worst case, 159\,\% in comparison to the status quo. This has important implications for operators of distribution networks, which must adapt their networks depending on the demand coincidence of new electricity technologies. 
\item We show how investments depend on network characteristics, such as size and consumer density. We find that the impact is more pronounced in large networks and in networks with spatially dispersed consumers, as frequently present in rural settlements. This means that technologically-induced changes in the coincidence factor require unequal infrastructure investments. Thus, investment budgeting and regulatory compensation schemes must be re-designed to incorporate regional characteristics. 
\end{enumerate}


\noindent This paper is organized as follows. In \cref{sec:related_work}, we provide a background on demand coincidence and electricity distribution network design. Informed by this, we develop our model for designing electricity distribution networks in \cref{sec:CAVC}. We present both exact and heuristic solution approaches to our problem in \cref{sec:opt}, which we then use to evaluate the impact of demand coincidence on network cost using both real-world and synthetic networks in \cref{sec:impact}. Finally, we discuss the implications of our findings and our contributions in \cref{sec:disc}. Proofs of all propositions, runtime analyses, and additional information on the experiments are provided as supplements to this paper.

\section{Background}
\label{sec:related_work}

We first review 
approaches to model demand coincidence of electricity consumption (\cref{sec:rw:load_coincidence}). Thereafter, we provide an overview of potential approaches to model demand coincidence in decision problems for designing electricity distribution networks (\cref{sec:rw:network_design}) and finally contrast the coincidence factor against varying load levels (\cref{sec:coincidence_factor_contrast}).

\subsection{Modeling Demand Coincidence in Electricity Consumption}
\label{sec:rw:load_coincidence}


Electricity distribution networks are critical infrastructures that are built to serve consumer electricity demand at all times. The networks are typically radial, and conventionally built for power to flow from a single source to the locations of consumer loads. Thus, each line on the network is sized to carry the maximum electric current that all downstream loads will demand at once, which is called the \emph{coincident demand}. In general, the coincident demand $D$ is less than the sum of the peak demands of each individual load $D_i$. For any group of loads, the \emph{coincidence factor} $\gamma$ is the ratio of the coincident peak to the sum of the individual peaks, \ie $\gamma = \frac{D}{\Sigma_{i \in N} D_i}$ \citep{Dickert.2010}.

It has long been observed empirically that the coincidence factor decreases as the number of consumers in a group increases, and models have been developed for use by distribution network planners that approximate demand coincidence as functions of the number of consumers $N$ (a review can be found in {\citet{Dickert.2010}).\footnote{\SingleSpacedXI\footnotesize Some literature captures the same phenomenon using diversity factors, which are the inverse of coincidence factors. The terminology, however, is not always consistent, and the two terms are often used interchangeably.} We denote these models as a function $\gamma(N)$ and observe the following basic properties exhibited: for one consumer, the coincidence factor is always 1, \ie $\gamma(1) = 1$. Furthermore, $\gamma$ is convex and monotonically decreasing in $N$; and, $\gamma$ approaches a limit for large $N$. In addition to the number of consumers, the coincidence factor depends on the types of electrical devices present. In general, consumers with high-powered devices that operate over long or similar periods of time such as electric vehicles and water heating are characterized by a larger demand coincidence because the consumption of these high-powered devices is more likely to overlap, and peak flows thus become additive \citep[cf.][]{Dickert.2010}.

Several works aim at estimating the coincidence factor from observational data \NEW{\citep[e.g.,][]{Boait.2015,bollerslev2021coincidence,Herman.1993,Konstantelos.2014, Richardson.2010,roberts2019characterisation,tran2021simulation,Widen.2010}}. For instance, \citet{Herman.1993} analyze load data to model the demand coincidence among residential consumers in a descriptive manner. Their approach has been widely picked up by researchers. More recent publications use prescriptive bottom-up models of simulated appliance usage to derive demand patterns of individual households and groups of households \NEW{\citep[e.g.,][]{bollerslev2021coincidence,Richardson.2010,tran2021simulation,Widen.2010}}. The spread of smart meters in recent years has made it possible to gather real-time data from thousands of households within the same electricity network \NEW{\citep{Konstantelos.2014, roberts2019characterisation}}, which allows for a more precise estimation of the coincidence factor. In this paper, we later use a model of coincidence, originally proposed by \citet{Rusck.1956}, that parameterizes $\gamma(N)$ as a function of the number of consumers $N$. An illustration is in \cref{fig:coincidence_factor_function}, showing how a limit $\gamma_{lim}$ is approached when $N$ becomes large.


\begin{figure}[!htbp]
    \centering
    \includegraphics[width = 0.4\textwidth]{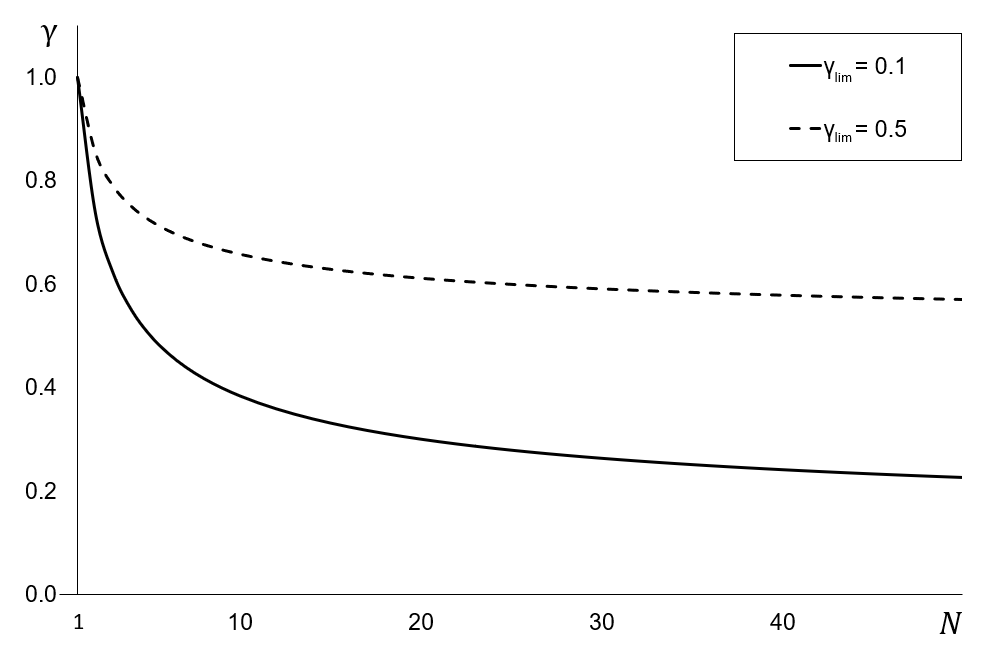}
    \caption{Example with two coincidence factors $\gamma(N)$ as a function of $N$ consumers.}
    \label{fig:coincidence_factor_function}
\end{figure}


Regarding applications in electricity networks, coincidence factors are, so far, mostly used to improve network operation (e.g., \citet{Boait.2015,Resch.2017}, \NEW{\citet{thormann2022estimation}}). \citet{Domingo.2011} consider a large network with three layers, namely a high voltage layer, a medium voltage layer, and a low voltage layer. Coincidence factors, however, are only applied at connections between these layers, \ie for the sizing of transformers. Within the low voltage network, the authors ignore coincidence factors, which leads to an inefficient over-dimensioning of network elements. Similarly, \citet{Parshall.2009} leverage the coincidence factor to estimate the demands of entire low-voltage distribution networks as a whole to help design the upstream parts of the electricity system. \citet{Kaur.2008} and \citet{Sauhats.} present models that take into account demand coincidence for the design of individual network elements. However, they use coincidence factors only to choose conductors (\ie line cross sections, conductor types, etc.) for a given network and do not regard effects on network layout. In contrast, it is unknown how demand coincidence affects the design of electricity distribution networks. \NEW{Our work contributes to the above literature by identifying the effect of demand coincidence on the entire distribution network, including network layouts as well as conductor sizes.}

In summary, there is considerable research on the effect of new technologies on the coincidence of electricity demand. However, there is a gap in research on the role of demand coincidence in electricity distribution networks. In particular, there is no approach for designing electricity distribution networks that takes into account the coincidence factor at the network line level. To overcome this shortcoming, we later develop a model for designing electricity distribution networks under variable degrees of demand coincidence. Our model considers demand coincidence separately for every vertex in the network and thus leads to a cost-effective network design. 

\subsection{Designing Electricity Distribution Networks}
\label{sec:rw:network_design}


Designing electricity distribution network aims to connect the individual consumers to the superordinate parts of the electricity system, such as the transmission network, in a cost-effective way. 
The design of distribution networks thereby needs to take into account three core constraints \citep[\eg][]{Weedy.2012}: 
\begin{enumerate}[(i)]
    \item Line sizing guarantees that the capacity of each network line is sufficient to supply the connected loads.
    \item A radial layout ensures that the flows of energy from the source to each load follow a unique path. \NEW{In practice, radial layouts have several advantages, such as easier operation, maintenance, and protection against faults \citep{short2004}, and are thus a common choice in the majority of related research \citep[see, \eg][]{vahidinasab2020overview}.}
    \item Voltage drops must be below a certain threshold. Along the lines of any electricity system, the voltage steadily drops due to the electrical resistance of the lines, leading to a lower voltage at the end-point of the system. Voltage drops beyond a certain threshold can damage the electric system and the connected electrical devices and, hence, must be avoided when designing a network. 
\end{enumerate}
In the following, we discuss related approaches for network design from different literature streams that consider all or a subset of the constraints~(i)--(iii).\footnote{\SingleSpacedXI\footnotesize In practice, the design process as a whole also involves sizing other components, such as circuit protection, reactive power compensation, monitoring and control systems, and local environmental constraints; in this paper, we focus on the core parts by addressing the network layout and line sizing only.} 


First, there is the literature on distribution network design (see \citet{Georgilakis.2015} and \NEW{\citet{resener2018optimization}} for an overview), where models exist that consider all three constraints. Prominent examples include the approaches by \citet{Boulaxis.2002, Cossi.2005, Falaghi.2011, Gan.2011, Kong.2009, Navarro.2009}, \NEW{as well as the group of works around reference network models \citep[\eg][]{Domingo.2011, miguez2002improved, mateo2018european}}. However, all solution approaches put restrictions on network layouts in order to reduce the solution space, \eg by restricting connections to vertices that are close in terms of Euclidean distance (\citet{Boulaxis.2002, Gan.2011}, \NEW{\citet{gholizadeh2016framework}}), close in terms of an existing street layout (\citet{Kong.2009, Navarro.2009}, \NEW{\citet{mehrtash2019graph, wang2022practical}}), or close in terms of pre-existing network infrastructure (\citet{Cossi.2005, Falaghi.2011}, \NEW{\citet{nahman2020radial}}). Therefore, the solution approaches disregard many potentially cost-effective layouts where line connections are long. Our model, which we develop later, is similar to the presented approaches that incorporate the constraints (i)--(iii); however, \NEW{unlike related solution approaches}, we do not restrict network layouts and show that introducing the coincidence factor leads to cost-effective network layouts that often rely on longer connections.


Second, the literature on distribution network reconfiguration \NEW{(see \cite{mishra2017comprehensive} for an overview)} is closely related, because the networks need to satisfy the same constraints (i)--(iii). However, this literature is focused more on operational planning: Instead of minimizing network \emph{investment} cost, network reconfiguration problems optimize for minimum \emph{operational} cost by minimizing power losses (e.g., \citet{Avella.2005}, \NEW{\citet{essallah2020optimization}}, \citet{Jabr.2012, Parada.2010}, \NEW{\citet{zhan2020switch}}). \NEW{Along with power losses, recent works have also considered multiple objectives, \eg to additionally improve load balancing \citep{ali2021novel,islam2019mitigating} or system reliability \citep{duan2015reconfiguration,paterakis2015multi,sultana2016review}.} \NEW{Regardless of the objective function,} the majority of reconfiguration approaches restrict the number of potential network layouts, because, in the short-term, only a limited number of edges can be reconfigured. Also, the capacity for any given connection is usually fixed in network reconfiguration problems to the line type present in the field. \REVTWO{These restrictions simplify the network design problem, but may lead to suboptimal solutions.} \NEW{We contribute to the literature by \emph{not} restricting the number of potential network layouts and by allowing a choice of several line types.}


\NEW{
Third, related to the previous streams of literature is research on distribution network expansion planning. Unlike the design and reconfiguration problems mentioned above, expansion planning also considers the temporal dimension of decisions. As such, expansion planning not only considers the decision around where to invest, but also when to do it (see \citet{vahidinasab2020overview} for a recent review). In these works, the objective function needs to consider both investment and operational costs (or, alternatively, their net present value). Along with the usual investment costs, operational costs are typically considered in the form of energy losses \citep{jabr2012polyhedral,ravadanegh2014optimal,yao2014multi} or reliability-related costs \citep{jooshaki2021enhanced, munoz2017distribution, tabares2022multistage}. In the latter literature stream, approaches use multi-objective functions that combine cost and reliability measures (a recent review is provided by \citet{aschidamini2022expansion}). Usually, the literature on expansion planning considers constraints (i) to (iii); however, unlike in our work, coincidence factors are \emph{not} included.}

In summary, while there are many models for designing electricity distribution networks, we are not aware of one that accounts for variable degrees of demand coincidence. As a remedy, we later develop our \pns network design model. It is based on the reconfiguration model by \citet{Avella.2005} but adapted to demand coincidence. In particular, we need to make multiple extensions in our \pns model in order to allow for multiple line types, optimizing for a different objective function, and, most importantly, integrating the coincidence factor. The integration of the coincidence factor is particularly challenging: Flow conservation is a key assumption of models for distribution network planning \citep[such as in][]{Avella.2005} but, in the context of variable demand coincidence, it no longer applies. Due to the coincidence factor, peak flows into a vertex can be smaller than the demand of this vertex and all outgoing flows combined. Our \pns model is able to take this into account.


\subsection{\NEW{Coincidence Factors Versus Varying Load Levels}}
\label{sec:coincidence_factor_contrast}

\NEW{In the context of electricity network planning, both coincidence factors and varying load levels refer to distinctively different concepts. As explained in \cref{sec:rw:load_coincidence}, the coincidence factor is the ratio of the coincident peak loads to the sum of the individual peak loads. The concept of varying load levels expresses that load exhibits different levels over time, such as over the course of a day or in different time-of-use periods \citep{gholizadeh2019electric}. Varying load levels do \emph{not} express how individual peak loads are related to the peak load in a system. Load levels are often required to consider \emph{operational aspects}, such as the reconfiguration of networks to reduce cost from power losses \citep{lee1988method,de2010optimal} as well as the placement of distributed generators to reduce losses \citep{esmaeili2015simultaneous}, the improvement of voltage profiles \citep{sajjadi2013simultaneous}, or the improvement of reliability \citep{ziari2012integrated}. In contrast, coincidence factors are used to model worst-case scenarios \citep{Dickert.2010} as required for \emph{long-term network design and investment decisions}. Therefore, designing electricity networks under variable degrees of demand coincidence should not be confused with network planning using different load levels.}

\section{The \pns Model} 
\label{sec:CAVC}

\subsection{Problem Statement}


Designing electricity distribution networks corresponds to the decision problem of connecting a given set of demand locations (\eg buildings) with a single source location (\ie the transformer to the superordinate network). Between the locations, network lines of different types can be built. Each line type has a specified capacity (\ie its cross section). The objective for the decision-maker is to minimize investment costs consisting of construction and material costs. The problem is subject to the following constraints.  
\begin{enumerate}[(i)]
\item\emph{Line sizing and demand coincidence.} The capacity of a network line must be large enough to support the electric current (\ie the flow). Note that, when choosing the capacity, we consider the fact that the peak demands of individual loads are partly coinciding. For every line in the network, we discount the flow by the line-specific coincidence factor. 
\item\emph{Radial layout.} Networks layouts must be radial, so that all energy flows from the source to each load follow a unique path.
\item\emph{Voltage drops.} The flow of electric current through a line causes a voltage drop. The voltage drop accumulates over consecutive lines. At any point in the network, it must remain below a threshold prescribed by industry norms.
\end{enumerate}

\subsection{Mathematical Formalization}

We now formalize the decision problem named the \emph{\pnl}~(\pns). An overview of the notation is provided in \Cref{tbl:notation}. The appropriate unit conversions and material constants for real-world settings can be found in \Cref{sec:add_info_case} in the supplements.

\begin{table}[ht]
\caption{Notation. \label{tbl:notation}}
{\resizebox{\columnwidth}{!}{%
\begin{tabular}{lll}
\toprule
\textbf{Symbol} & \textbf{Description} & \textbf{Unit/range} \\
\hline
$ G											$& Directed multigraph & $G = (V, E)$ \\
$ V											$& Set of all vertices & \\
$ N											$& Number of vertices & $N = |V|$\\
$ i, j									$& Indices of vertices & $i, j = 0, \ldots, N-1$ \\
$ D_i										$& Demand of vertex $i$ & $D_0 = 0, D_{i \neq 0} = D > 0$. \\
$ E											$& Set of all directed edges & \\
$ k											$& Index of line type  & $k = 1, \ldots, |A|$  \\
$(i,j)^k								$& Directed edge from $i$ to $j$; $k$ denoting its type & $(i,j)^k \in E$ \\
$ A         						$& Set of edge capacities depending on line type $k$ (in ascending order)  &  \\
$ a_{ij}^k							$& Edge capacity & $a_{ij}^k > 0, \, a_{ij}^k \in A$ \\
$ C         						$& Set containing the maximum allowed flows depending on line type $k$ (in ascending order)  &  \\
$ c_{ij}^k							$& Maximum allowed flow & $c_{ij}^k > 0, \, c_{ij}^k \in C$ \\
$ \mathcal{N}_j					$& Set of all vertices that can be reached from $j$ & \\
$ \Gamma								$& Graph representing one solution of the problem & $\Gamma = (V, E'),$ with $E' = \left\{ (i,j)^k \in E \, \mid \, x_{ij}^k = 1 \right\}$ \\
$ \Gamma_i							$& Subgraph of $\Gamma$ including $j$ and all edges and vertices reachable from $j$ & \\
$ |\Gamma_j|						$& Number of vertices in $\Gamma_j$ & \\
$ \gamma(|\Gamma_j|)		$& Coincidence factor (discount factor depending on the number of vertices) & $ 0 < \gamma(|\Gamma_j|) \leq 1$ \\
$ \overline{D}_j				$& Undiscounted sum of all demands in $\Gamma_j$ & \\
$ d(i)									$& Depth of vertex $i$, \ie number of hops to reach $i$ from the source in $\Gamma$ \\
$ F_{ij}          			$& Flow through edge $(i, j)^k$ &  $F_{ij} > 0 $  \\	
$ l_{ij}								$& Length of edge $(i, j)^k$ &  $l_{ij} \in \mathbb{R}^+$ \\
$ P											$& Set of all paths from source vertex $0$ to any leaf vertex   & Set of edge sequences \\
$ p											$& Specific path from source vertex $0$ to a leaf vertex & Edge sequence, $p \in P$ \\
$ c_{\mathrm{c}}     		$& Construction costs & Monetary unit per distance \\
$ c_{\mathrm{m}}   			$& Material costs & Monetary unit per distance per capacity unit \\
$ U_i										$& Voltage at vertex $i$ & $U_i > 0$ \\
$ U											$& Voltage at transformer & $U > 0, \, U_0 = U$ \\
$ U_{\mathrm{crit}}			$& Critical voltage level & $U_{\mathrm{crit}} > 0$ \\
$ Q											$& Threshold value for voltage drop & $Q = U - U_{\mathrm{crit}}$ \\
$ x_{ij}^k     					$& Decision variable for edge from vertex $i$ to $j$ with capacity $a_{ij}^k$ & $x_{ij}^k \in \left\{0, 1\right\} $\\
\bottomrule
\end{tabular}}
}
{}
\end{table}

Let $G = (V, E)$ denote a complete directed multigraph without loops. The set of vertices $V = \left\{ 0, \ldots, \abs{N - 1} \right\}$ represents locations. Each vertex $i \in V$ has a given demand $D_i$. The source location (\ie the transformer) is defined as vertex $0$, and we further set $D_0 \define 0$. The set of edges $E$ contains all potential network lines. Each edge $(i, j)^k \in E$ from vertex $i$ to vertex $j$ has a discrete capacity $a_{ij}^k \in A$ (\ie a discrete cross section). The index $k$ indicates the line type. The maximum allowed flow (\ie electric current) for any given line type $k$ is given by the flow capacities $c_{ij}^k \in C$. Furthermore, each edge $(i, j)^k$ has a given length $l_{ij}$ that is independent of its type (\ie the same for all $k$). The decision variable $x_{ij}^k \in \left\{ 0, 1 \right\}$ indicates whether an edge of type $k$ from vertex $i$ to vertex $j$ should be built. We denote the subgraph representing one solution of our problem $\Gamma(x) = (V, E') \, \text{as} \, E' = \left\{ (i,j)^k \in E \, \mid \, x_{ij}^k = 1 \right\}$. Furthermore, let $\Gamma_j$ denote the subgraph of $\Gamma$ encompassing a certain vertex $j$ and all vertices and edges that can be reached from $j$ in the direction of the flow. The number of vertices in $\Gamma_j$ is denoted $|\Gamma_j|$. 

The \pns is then given by
\begin{align}
	& \text{min} && \sum\limits_{(i, j)^k \in E} x_{ij}^k \, [l_{ij} c_{\mathrm{c}} + l_{ij} c_{\mathrm{m}}\, a_{ij}^k] 	\label{eq:obj_fun}	
	\\
  & \text{s. t.} && \sum \limits_{k \in \{1, \ldots, |C|\}} x_{ij}^k c_{ij}^{k} \geq  F_{ij} \, ,	\quad &&\forall (i, j) \in E \, ,   \label{eq:current}
	\\
	& &&  \sum\limits_{j} F_{ji} - \sum\limits_{j} F_{ij} = \gamma \left(\abs{\Gamma_i} \right) \, D_{i} \, -  \nonumber
	\\
	\newsubeqblock
	\mysubeq & && \quad \sum\limits_{j} \sum \limits_k x_{ij}^k \left( \left[ \gamma \left( \abs{\Gamma_j} \right) - \gamma \left(\abs{\Gamma_i} \right) \right] \overline{D}_j \right) \, ,	\quad &&\forall i \in V \setminus \{ 0 \}	 \, ,  \label{eq:flow} 
	\\
	\mysubeq & && \abs{\Gamma_i} = 1 + \sum \limits_{j} \sum \limits_{k} x_{ij}^k \abs{\Gamma_j} \, , \quad &&\forall i \in V	\, ,  \label{eq:gamma_i}
	\\
	\mysubeq & && \overline{D}_i = D_i + \sum \limits_{j} \sum \limits_{k} x_{ij}^k \, \overline{D}_j \, , \quad &&\forall i \in V	\, ,  \label{eq:d_bar_i}
	\\
	\mysubeq & && \abs{\Gamma_0} = N - 1 \, ,   \label{eq:gamma_0}
	\\
	\mysubeq & && \overline{D}_0 = \sum \limits_{j} D_j \, ,   \label{eq:d_bar_0}
	\\
	\mysubeq & && \sum\limits_{i} \sum \limits_k  x_{ij}^k = 1 \, , 	\quad &&\forall j \in V \setminus \{ 0 \}	\, ,  \label{eq:tree}
	\\
	\newsubeqblock
	\mysubeq & && \sum\limits_{k} x_{ij}^k \frac{a_{ij}^k}{l_{ij}} (U_i - U_j) = F_{ij} \, , \quad  &&\forall (i, j) \in E \, ,     \label{eq:constr_voltage_1}
	\\
	\mysubeq & && U_i \geq U_{\mathrm{crit}} \, , \quad &&\forall  i \in V \, ,     \label{eq:constr_voltage_2}
		\\
	\mysubeq & && U_0 = U  \, ,     \label{eq:constr_voltage_3}
 \\
   &  && \NEW{x_{ij}^k \in \{0,1\} \, ,}	\quad && \NEW{\forall (i, j)^k \in E \, .}   \label{eq:binary} 
\end{align}


\noindent
The objective in \Cref{eq:obj_fun} is to minimize investment costs. If an edge with capacity $a_{ij}^k$ from vertex $i$ to vertex $j$ is built, construction costs of $l_{ij} c_{\mathrm{c}}$ are incurred (depending only on the length of the edge) and material costs of $l_{ij} c_{\mathrm{m}}\, a_{ij}^k$ are incurred (depending on the length and the capacity). The line sizing constraint in \Cref{eq:current} requires the edge $(i, j)^k$ to be sufficiently large to support the peak flow $F_{ij}$. Thus, the peak flow $F_{ij}$ must not be larger than the capacities $c_{ij}^k$.

\Crefrange{eq:flow}{eq:tree} define the flows and ensure the radial layout (including connectivity) of the network. The peak flows $F_{ij}$ are defined recursively in \Cref{eq:flow}. In the special case of a uniform coincidence factor of $\gamma \equiv 1$, \Cref{eq:flow} simplifies to $\sum\limits_{j} F_{ji} - \sum\limits_{j} F_{ij} = D_{i}$, and, as a result, \Cref{eq:gamma_i,eq:d_bar_i} would no longer be needed. In this special case, all flows into a vertex minus all flows out of this vertex equal the demand of the vertex. Except for this special case, the peak flows in the \pns are not conserved---as they do not fully coincide. To take this non-coincidence of electricity demand into account, the correction term at the end of \Cref{eq:flow} is needed. This correction term considers the direct neighbors of vertex $i$, \ie all vertices $j$ with $\sum \limits_{k} x_{ij}^k = 1$. Each neighbor $j$ connects a subgraph $\Gamma_j$ to vertex $i$. The number of vertices in a subgraph $|\Gamma_j|$ determines the magnitude of the coincidence factor $\gamma(|\Gamma_j|)$ and thus the flows going out of vertex $i$. The correction term determines the discount of outgoing flows relative to incoming flows, which is given by the difference in the coincidence factors $\gamma(|\Gamma_j|)$ and $\gamma(|\Gamma_i|)$.

\NEW{
We illustrate \cref{eq:flow} using a small example network in \cref{fig:example_flow_equation}. The network consists of transformer at vertex 0, vertices 1 and 2 with peak demands $D_1$ and $D_2$, as well as two lines (0,1) and (1,2). (For further simplification, we assume only one line type to be available and can thus omit index $k$.) We now formulate \cref{eq:flow} for the central vertex, \ie vertex $i=1$. Inflows $F_{01}$ over edge $(0,1)$ need to serve the peak demands $D_1$ and $D_2$. Since there are \emph{two} vertices that need to be served downstream by the subgraph $\Gamma_1$ (since $\Gamma_1$ contains vertices 1 and 2), the sum of these peak demands is discounted with $\gamma(2)$, \ie $F_{01}= \gamma(2) (D_1 + D_2)$. Analogously, outflows $F_{12}$ over edge $(1,2)$ consist of the peak demand $D_2$, that is discounted by $\gamma(1)$, \ie $F_{12} = \gamma(1)D_2$. Taking the difference between inflows and outflows thus yields $F_{01}-F_{12} = \gamma(2) (D_1 + D_2) - \gamma(1)D_2$. Rearranging the right-hand side of the previous equation leads to $F_{01}-F_{12} = \gamma(2) D_1 - (\gamma(1) - \gamma(2))D_2$, which is \cref{eq:flow} for vertex 1 in this example network. The first term of the right-hand side,  $\gamma(2) D_1$, corresponds to the discounted demand of vertex 1. The second term, $(\gamma(2) - \gamma(1))D_2$, is the correction term that accounts for the fact that the demand of vertex 2 is discounted more strongly on line (0,1) than on line (1,2).}

\begin{figure}
    \centering
    \includegraphics[width=0.5\textwidth]{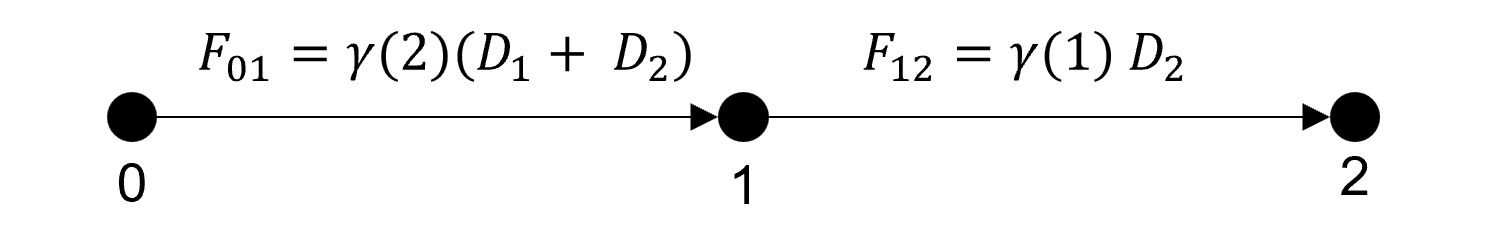}
    \caption{\NEW{Flows in an exemplary solution of the \pns for an instance with three vertices.}}
    \label{fig:example_flow_equation}
\end{figure}

\Cref{eq:gamma_i,eq:d_bar_i} define $|\Gamma_i|$ and $\overline{D}_i$ recursively: The number of vertices inside a subgraph $|\Gamma_i|$ always equals one plus the sum of all subgraphs downstream. The term $\overline{D}_i$ denotes the (undiscounted) sum of demands downstream a certain vertex $i$. This is equal to $D_i$ plus the sum of all demands downstream. \Cref{eq:gamma_0,eq:d_bar_0} define vertex $0$ as the source by setting the values for $|\Gamma_0|$ and $\overline{D}_0$ to the overall number of loads and the total demand of all loads, respectively. 
\Cref{eq:tree} ensures that every vertex (except the source) is entered by exactly one edge.\footnote{\NEW{In rare situations, two cables of an identical type may be installed in parallel between the same two vertices in order to cover exceptionally large demands (such as for large industrial and commercial consumers). We refrain from permitting this in our model; however, such an extension could be achieved without further changes to the model by including a large fictitious cable type with the equivalent electrical characteristics of two smaller, real-world cable types that are connected in parallel.}}

\Crefrange{eq:constr_voltage_1}{eq:constr_voltage_3} limit the magnitude of voltage drops. Specifically, \Cref{eq:constr_voltage_1} follows Ohm's law and describes the voltage drop between two connected vertices $i$ and $j$.  According to Ohm's law, the voltage drop is proportional to the peak flow (\ie the discounted sum of the peak demands). It is also proportional to the edge length $l_{ij}$ and inversely proportional to the line's cross section (\ie the edge capacity) $a_{ij}^k$. Modeling power flow in this way is valid under certain assumptions (\eg the demand at a node must be independent of the node voltage, and the inductance and capacitance of both lines and loads must be negligible), and these assumptions are typically made for electricity distribution grids as they provide a useful approximation for network planning purposes. \Cref{eq:constr_voltage_2} demands that the voltage $U_i$ of any vertex $i$ cannot drop below a critical voltage level $U_{\mathrm{crit}}$. \Cref{eq:constr_voltage_3} sets the voltage at the source to the nominal level $U$.

\subsection{Complexity} 
\label{sec:complex}

The \pns is NP-hard. In the supplements to this paper, we derive NP-hardness by reduction: we show that the \pns is a generalized form of the problem presented in \cite{Brimberg.2003}, which is known to be NP-hard. Specifically, the problem in \citet{Brimberg.2003} results from the \pns by assuming a uniform coincidence factor $\gamma \equiv 1$ and by setting $U_\textrm{crit} = 0$. 



Furthermore, we note that \Crefrange{eq:obj_fun}{eq:constr_voltage_3} give a mixed-integer nonlinear program. More precisely, nonlinearities are found in \Crefrange{eq:flow}{eq:d_bar_i}, as well as \Cref{eq:constr_voltage_1}. The quadratic nonlinearities in \Cref{eq:gamma_i}, (\ref{eq:d_bar_i}), and (\ref{eq:constr_voltage_1}) can be resolved by using the Big~M method. Linearization of \Cref{eq:flow} is more complex. The equation contains a product of the decision variable $x_{ij}^k$ with the auxiliary decision variable $\overline{D}_j$ and with the nonlinear function $\gamma(|\Gamma_j|)$, which depends on the auxiliary decision variable $|\Gamma_j|$. Therefore, the linearization requires first a piecewise linearization of the coincidence factor $\gamma(|\Gamma_j|)$, after which the equation still contains a cubic nonlinearity. All linearizations are provided in \Cref{sec:lin} in the supplements. Nevertheless, such linearization is not helpful for our problem. As we show later, the complexity of our problem makes it intractable to use common commercial mixed-integer programming (MIP) solvers even for relatively small networks. For this reason, we later develop heuristics as solution approaches. 


\subsection{Solution Properties} 
\label{sec:prop}

There are four properties of the optimal solution of the \pns that we later use for developing our solution procedures. The first two properties describe the network layout of the optimal solution, while properties three and four characterize the capacities.

\subsubsection{Optimal Network Layout for Low Demand Settings.}

The minimum spanning tree~(MST) is the cycle-free network connecting all vertices with the shortest total edge length \citep{Prim.1957}. Let $\Gamma^{\mathrm{MST}}$ denote a solution of the \pns with an MST layout. 

\vspace{0.15cm}
\begin{remark} \label{pro:min}
\textit{For $D_i \to 0$ (\REVTWO{or, alternatively, $D_i > 0$ and  $a_{\mathrm{min}} = \underset{k}{\min} \; a_{ij}^k$ satisfying the flow and voltage drop constraints}}), the MST layout is the optimal solution to the \pns.
\end{remark}

\subsubsection{Optimal Network Layout for High Demand Settings.}

In settings with sufficiently high demand, many network layouts will not yield feasible solutions, even when all capacities are set to the maximum possible value. Let $\Gamma^{\mathrm{Star}}$ (``star network'') denote a solution of the \pns in which every vertex is connected directly to the source $0$, \ie $x_{ij} = 1$ for $i = 0$ and, otherwise, $x_{ij} = 0$. 

\vspace{0.15cm} 
\begin{remark} \label{pro:star}
\textit{If feasible solutions to the \pns exist, the star network is one of these solutions.}
\end{remark}

\subsubsection{Line Capacity Ratios.} 

For a given radial network layout, the remaining decisions are which capacities to assign to the connections of this layout.  To later develop efficient heuristics for optimizing the line capacities (\Cref{sec:p2}), we now derive useful properties regarding the capacities, which must hold true in order to minimize costs. For ease of notation, we omit the superscript $k$ and treat the capacities $a_{ij}$ as a decision variable. Furthermore, we use a continuous relaxation of the problem, \ie we allow the choice of continuous capacities ($A = \mathbb{R}^+$). As we see later, the following \cref{pro:ratio} proves to be very powerful in situations where the constraint for the voltage drops is binding. This is typically the case in real-world applications. 

\vspace{0.15cm}
\begin{proposition}[Line capacity ratios] \label{pro:ratio}
For any given network layout and for continuous capacities (\ie $A = \mathbb{R}^+$), the capacities that minimize the cost fulfill
	\begin{align}
		\quad \frac{a_{ij}}{a_{mn}} = \sqrt{\frac{F_{ij}}{F_{mn}}} 
		\label{eq:branch_ratio}
	\end{align}
\REVTWO{for all combinations of edges} $(i,j)$ and $(m,n)$ in the same path $p \in P$ and if \Cref{eq:constr_voltage_2} is binding.
\end{proposition}

\NEW{To convey the intuition behind the square-root relation in \Cref{pro:ratio}, we make use of two statements that follow immediately from \cref{eq:branch_ratio}: (1)~Optimal line capacities increase with flows. (2)~This increase is subproportionate, \ie the ratio of line capacities grows slower than the flows (according to the square root on the right-hand side). The intuition behind these statements can be understood by analyzing the voltage drop occurring over a line segment $\Delta U \define U_r - U_s = \frac{l_{rs}}{a_{rs}} F_{rs}$. We observe that the voltage drop $\Delta U$ decreases with the increase of $a$. By taking the partial derivative of the voltage drop $\frac{\partial \Delta U}{\partial a_{rs}} = - \frac{l_{rs}}{a_{rs}^2} F_{rs} $, we can further observe that the decrease in voltage drop $\Delta U$ from adding line capacity is more pronounced the larger the flows $F_{rs}$. This means, for lines with larger flows, there is a higher benefit of increasing line capacities in comparison to lines with smaller flows. This explains our first statement~(1). To further understand the intuition behind the subproportionality in the second statement~(2), we observe that the partial derivative is quadratically approaching zero for large values of $a_{rs}$. This means that the larger the capacities, the smaller the effect of an additional increase of $a_{rs}$. This diminishing effect explains the square-root-relation in \Cref{pro:ratio}. In sum, it follows that one should counteract larger flows using larger line capacities; however, the marginal benefit of adding line capacities decreases with the square root. 
}

\subsubsection{Decreasing line capacities.} 

The following corollary states that, starting from the source to the leaves within the network, line capacities are monotonically decreasing. 
\begin{corollary} \label{pro:decrease}    
\textit{Under the above assumptions, the cost-minimizing line capacities for any given network layout decrease when moving downstream.}
\end{corollary}
\Cref{pro:decrease} follows directly from \Cref{pro:ratio}. Formally, the flow $F_{ij}$ through an edge $(i,j)^k$ is the sum of all demands downstream to this edge. All demands are positive real numbers, \ie $D_i \in \mathbb{R}^+$, for all $i$. Therefore, the flows decrease when moving downstream.

\NEW{Finally, while \cref{pro:ratio} and \cref{pro:decrease} later help us to design effective solution methods, it remains to note that both properties only allow conclusion to be made about line capacities within the same path. (This is because \cref{eq:constr_voltage_1} modeling the voltage drop only relates edges to each other that belong to the same path). Moreover, under the assumption of continuous line capacities, it would suffice to consider \REVTWO{each line segment in the evaluation of \cref{pro:ratio} only once.} However, since real-world line capacities are discrete, deviations from the optimal ratio in \cref{pro:ratio} occur frequently in practical applications. Therefore, in practice, the deviations require iterative and repeated evaluations of the ratio in \cref{eq:branch_ratio} for each line segment. \cref{pro:ratio} and \cref{pro:decrease} inform our solution approaches in the next section.}

\section{Optimization Methods} 
\label{sec:opt}

We now develop solution algorithms for our \pns. For this purpose, we divide the \pns into two sub-problems: (A)~generating the network layout and (B)~capacity optimization. 
Generating the network layout (A) is addressed by two sets of heuristics. These first create an initial solution (\cref{sec:p1}) and, given an initial solution as input, then make local improvements to the network layout (\cref{sec:p3}). To determine the feasibility and cost of the solutions, the algorithms rely upon additional input in the form of line capacities $\{a_{ij}\}$. These are determined in the second sub-problem, \ie capacity optimization~(B), which is presented later in \Cref{sec:p2}.
For a given layout, capacity optimization determines suitable line types for each edge in the network. 
We conclude the section by benchmarking the performance of our solution algorithms in terms of solution quality and runtime against alternative approaches (\cref{sec:comp}).

\subsection{Generating the Initial Network Layout} 
\label{sec:p1} 

The objective of generating an initial network is to create a network layout but without considering different line types. To this end, we reduce the multigraph $G$ to a graph $\widetilde{G} = (V, \widetilde{E})$ whereby multiple edges between the same vertices are replaced by a single edge, that is, by setting $A = \{ a \}$. In the optimization problem in \Crefrange{eq:obj_fun}{eq:constr_voltage_3}, this corresponds to dropping the index $k$ so that the decision variable becomes $x_{ij} \in \{0, 1 \}$. The decision variable $x_{ij}$ then indicates whether an edge $(i, j) \in \widetilde{E}$ from vertex $i$ to vertex $j$ should be built. The resulting network layout is represented by $\{ x_{ij} \}$. To generate such initial layouts, we use a combination of the minimum spanning tree~(MST) and the Esau-Williams algorithm, as described in the following. 

\subsubsection{Minimum Spanning Tree.} \label{sec:MST}
The MST connects all vertices, so that the total length of all edges is minimized. In our implementation of the MST, we rely on Prim's algorithm \citep{Prim.1957}, which has a runtime of $O((N-1)\log\,N$).

In \Cref{pro:min}, we showed that the MST is the optimal solution to the \pns for situations in which voltage drops can be neglected. This makes the MST favorable for instances with low demands $D_i$. However, there might be situations where the MST yields an infeasible layout. This occurs mainly in situations with high demand and coincidence, which lead to a violation of the voltage drop constraints (\Crefrange{eq:constr_voltage_1}{eq:constr_voltage_3}), even if the largest possible line capacities are chosen later during capacity optimization. Here, another starting layout must be found. In these situations, we use the Esau-Williams algorithm to arrive at a feasible starting point. 

\subsubsection{Optimization Using the Esau-Williams Algorithm.}

We use the MST as a starting network layout whenever possible (\ie if the MST is a feasible layout). However, if the MST yields an infeasible layout, we revert to other layouts with an increased branching. To generate these networks with increased branching, the Esau-Williams (EW) algorithm \citep{Esau.1966} is utilized. 

The EW algorithm is a greedy procedure that provides near-optimal solutions to the capacitated minumum spanning tree (CMST) problem \citep{Bruno.2002}. The CMST problem 
aims to find the cycle-free network connecting all $N$ vertices with the shortest total edge length. Thereby, each of the subtrees directly connected the source must contain at most $K$ vertices. For $K = 1$, the Esau-Williams algorithm always returns the star network. For $K = N - 1$, the solution is close to the MST. 
The runtime of the EW algorithm is in $O(N^2 \log N)$ and is known as a highly efficient algorithm for solving the CMST problem \citep{Jothi.2004}.

The overall optimization approach for determining initial layouts is described in \cref{alg:approach}. The EW~algorithm is used if the MST algorithm produces an infeasible layout. In such cases, the EW algorithm is used to find the network layout for $K = \tfrac{N}{2}$. If this is still infeasible, the EW algorithm is used with $K = \tfrac{N}{4}$, $K = \tfrac{N}{8}$, \etc, until a feasible layout is found. The feasibility is evaluated using routines for capacity optimization (\ie \textsc{CapacityOptimizationMethod}), which are described later in \cref{sec:p2}. The \textsc{CapacityOptimizationMethod} returns a suggested set of capacities $\{a_{ij}\}$ for a given layout $\{x_{ij}\}$. Essentially, the iterative procedure allows us to tune the branching of the starting layout (\ie the number of vertices connected to the source) depending on the demand situation. This approach to determine the initial layout is exemplified in \Cref{fig:opt_scheme}. As the EW algorithm eventually returns the star network, we can guarantee that a feasible solution will be found if it exists (\Cref{pro:star}).

\begin{algorithm} [htbp]
    \SingleSpacedXI
	\tiny 
	\caption{Generating the Initial Layout} 															\label{alg:approach}
	\begin{algorithmic}[1]
		\State Create MST by using $\textsc{MinimumSpanningTreeAlgorithm}$
		\State $\{ a_{ij} \} \gets \textsc{CapacityOptimizationMethod}(\{x_{ij}\})$	
		\State $n \gets 1$
		\While{network is infeasible}
		\State Perform $\textsc{EsauWilliamsAlgorithm}(K \gets \ceil{\tfrac{N}{2^n}})$    \label{line:EW_round}
		\State $\{ a_{ij} \} \gets \textsc{CapacityOptimizationMethod}(\{x_{ij}\})$	
		\State $n \gets n + 1$
		\EndWhile
		\State \textbf{return} $X^*$ 	
	\end{algorithmic}	
\end{algorithm}

\begin{figure}[htbp]
\begin{center}
\includegraphics[width = .6\textwidth]{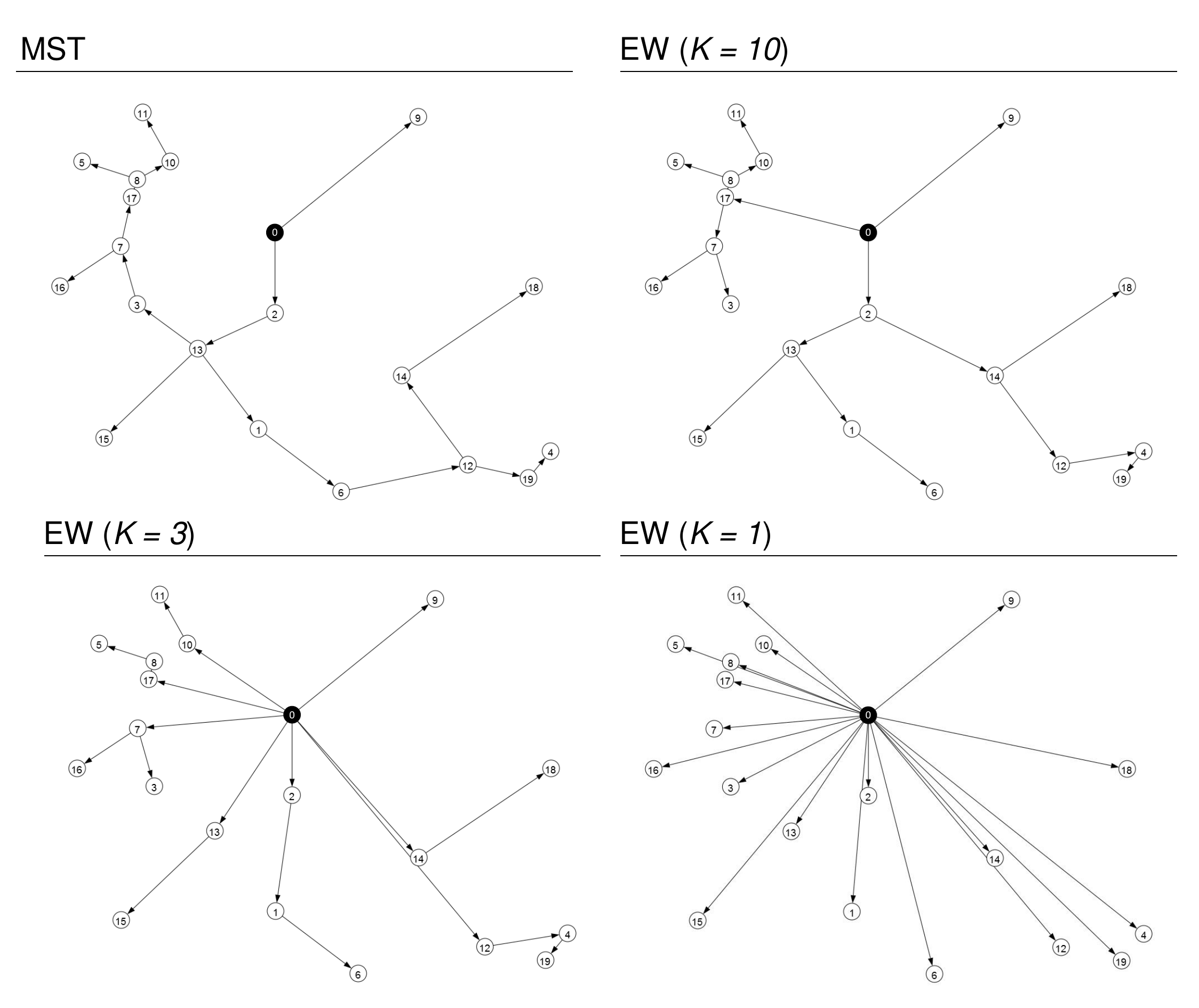}
\end{center}
  \SingleSpacedXI
	\footnotesize 
\caption[Approach to determine initial layout]{\raggedright Initial layouts generated for an example network by the MST algorithm (top left). The MST returns the network with the lowest branching. Afterward, the Esau-Williams algorithm is applied with decreasing values for $K$. For the Esau-Williams algorithm, the branching increases with smaller values for $K$. In the case of $K=1$, the star network is returned (bottom right). The search process is stopped once a feasible solution is found. }
\label{fig:opt_scheme}
\end{figure}




\subsection{Improving an Existing Network Layout} 
\label{sec:p3} 

Departing from an initial layout, we develop a Tabu Search metaheuristic to improve the layout. The algorithm takes a given layout $\{x_{ij}\}$ as input and then subsequently modifies it to improve cost. The Tabu Search metaheuristic thereby uses a short-term memory in the form of a tabu list to find solutions more efficiently \citep{Gendreau.2014,Glover.1989}. Our implementation of Tabu Search uses a tabu list $T$ of fixed length $l_\mathrm{T}$. The list contains both recently deleted edges to avoid cycling back to previous solutions and unfavorable edges, \ie edges that were recently explored without leading to improvements (note that we also experimented with two separate tabu lists but this did not yield better results). 
	\begin{algorithm}
		\SingleSpacedXI
		\tiny 
		\caption{Tabu Search} 																																																	\label{alg:tabu}
		\textbf{Input:} Network layout $\{x_{ij}\}$		
		\begin{algorithmic}[1]
			\State $\{ a_{ij} \} \gets \textsc{CapacityOptimizationMethod}(\{x_{ij}\})$																						\label{line:tabu_1}
			\State $c^* \gets \textsc{Cost}\left(\{x_{ij}\}, \{a_{ij}\}\right)$	 	\Comment{Determine initial cost}			\label{line:tabu_2}
			\State $X^* \gets \{x_{ij}\} $ 				 																																						\label{line:initial_tabu}
			\State $T \gets \emptyset$ 																							\Comment{Initialize tabu list}					   \label{line:tabu_list}
			\For{$s \in \{ 1, \ldots, s_{\mathrm{Tabu}} \}$}																																				\label{line:tabu_for}
				\State Select random disabled edge $(i,j) \not \in T$, such that $x_{ij} = x_{ji} = 0 $															\label{line:tabu_select}
				\State $x_{ij} \gets 1$   																					\Comment{Add edge $(i,j)$}											\label{line:add_tabu}
				\State $t \gets (i,j)$																							\Comment{$(i,j)$ is the candidate for the tabu list}  \label{line:tabu_cand}
				\State Compute list $C$ containing all edges comprising the cycle																							\label{line:tabu_cycle}
				\For{each edge $(p,q) \neq (i,j) \in C$}																																						\label{line:tabu_loop_for}
					\State $x_{pq} \gets 0$																						\Comment{Delete edge inside the cycle to create radial layout}				\label{line:tabu_delete}
					\State Check direction of all edges $(m,n) \in C$ and reverse if necessary   																							\label{line:tabu_direction}
					\State $\{ a_{ij} \} \gets \textsc{CapacityOptimizationMethod}(\{x_{ij}\})$																				\label{line:tabu_peca}
					\State $c \gets \textsc{Cost}\left(\{x_{ij}\}, \{a_{ij}\}\right)$ \Comment{Determine new cost}										\label{line:tabu_cost}
					\If{$c < c^*$}
						\State $c^* \gets c$ and $X^* \gets \{x_{ij}\}$									\Comment{Update network if cost is cheaper}									\label{line:tabu_if}	
						\State $t \gets (p,q)$																				 	\Comment{$(p,q)$ is the new candidate for the tabu list}  \label{line:tabu_cand_2}
					\EndIf
					\State $x_{pq} \gets 1$																						\Comment{Close cycle to reset old configuration}						\label{line:tabu_reverse_1}
				\EndFor
				\State $\{x_{ij}\} \gets X^*$																																												\label{line:tabu_reverse_2}
				\State $T \gets T \cup t$ 																					\Comment{Update tabu list}				\label{line:tabu_update_start}
				\If {$\abs{T} > l_\mathrm{T}$} remove first element in $T$																										\label{line:tabu_update_end}
				\EndIf
			\EndFor
		\State \textbf{return} $X^*$ 																																														\label{line:return_tabu}
		\end{algorithmic}	
	\end{algorithm}

The pseudocode of our Tabu Search algorithm is provided in \Cref{alg:tabu}. In \crefrange{line:tabu_1}{line:initial_tabu}, the heuristic determines capacities and initial cost. In \cref{line:tabu_list}, the tabu list $T$ is initialized by an empty list. The main part of the algorithm starts in \cref{line:tabu_for} and is executed for a pre-defined number of $s_{\mathrm{Tabu}}$ iterations. The algorithm adds a random edge $(i,j)$ in \crefrange{line:tabu_select}{line:add_tabu}. This edge must not be in the tabu list $T$. It is then saved as a potential candidate for the tabu list in \cref{line:tabu_cand}. This means that if no better solution is found during the following process, $(i,j)$ should not be added for the next iterations because it is unfavorable. 
The added edge creates a cycle $C$, which is determined in \cref{line:tabu_cycle}. In \cref{line:tabu_loop_for}, the algorithm iterates over each $(p,q) \in C$ that is not the newly added edge. The edge $(p,q)$ is deleted in \cref{line:tabu_delete} to obtain a radial layout. This may require the direction of some edges to be reversed in \cref{line:tabu_direction}. The capacities $\{ a_{ij} \}$ and cost $c_s$ for the new network layout are determined in \crefrange{line:tabu_peca}{line:tabu_cost}. If the resulting costs are lower than the current best solution, the algorithm saves the current network as the new best solution in \cref{line:tabu_if}. Additionally, the edge $(p,q)$ is then saved in the tabu list in \cref{line:tabu_cand_2} to avoid cycling back to the previous solution. In \crefrange{line:tabu_reverse_1}{line:tabu_update_end}, the changes are reversed before proceeding with the next iterations, and the tabu list is updated. Finally, the heuristic returns the network with the lowest costs in \cref{line:return_tabu}.

\subsection{Optimizing the Capacities} 
\label{sec:p2}

We now proceed with capacity optimization, which takes a network layout $\{x_{ij}\}$ as input and determines the corresponding line types $k$. \REVTWO{Early approaches for solving this problem (also known as conductor size selection~(CSS) problem) were based on dynamic programming \citep{ponnavaikko1982approach}, albeit only for feeders that do not exhibit any branching. For problem instances that include branched feeders and feeders serving a large number of consumers (as in this paper), later works \citep{Kaur.2008, wang2000practical, tram1988optimal} find that dynamic programming formulations lead to prohibitively large state spaces and computational demand. Here, integer programming-based methods \citep[\eg][]{franco2012optimal} have proven superior.} 

Accordingly, in \cref{sec:cap_problem_formalization}, we present an integer programming~(IP) formulation of the capacity optimization problem. However, as we show later in \Cref{sec:cap_opt_eval}), the solution using IP solvers is still impractical for larger problem instances because of excessive runtime. Therefore, we develop an additional solution heuristic in \Cref{sec:peca}.

\subsubsection{Problem Formalization.}\label{sec:cap_problem_formalization}

We now formalize the capacity optimization problem as an integer program. The main difference to the full \pns (\ie the combination of the two sub-problems) is the fact that capacity optimization only considers a subset of edges $E' \subset E$, namely these edges where $x_{ij} = 1$ has been determined in the first sub-problem. As a consequence, the flows are now given and $F_{ij}$ is no longer an auxiliary decision variable. This reduces the complexity of the problem, so that capacity optimization can be formulated as a binary integer program:
\begin{align}
	& \text{min} && \sum\limits_{(i, j)^k \in E'} x_{ij}^k l_{ij} \, a_{ij}^k															\label{eq:new_objective}
	\\
	&\text{subject to} &&  \sum \limits_{k \in \{1, \ldots, |C|\}} x_{ij}^k c_{ij}^k \geq  F_{ij} \, ,										&& \forall i, j \in \{0, \ldots, N - 1 \} \, , 				\label{eq:new_current}
	\\
	& && \sum \limits_k  x_{ij}^k = 1 \, , 																																								&& \forall (i,j)^k \in E' \, , 							 \label{eq:new_tree}
	\\
	& && \sum\limits_{(i, j)^k \in p} x_{ij}^k l_{ij} \frac{F_{ij}}{a_{ij}^k} \leq U - U_{\mathrm{crit}} = Q \, , 				&& \forall p \in P \, , 																				\label{eq:new_voltage}
    \\
	& && \NEW{x_{ij}^k \in \{0,1\}}, 				&& \NEW{\forall (i,j)^k \in E' \,.} 																				\label{eq:cap_binary}
\end{align}
\noindent
Since the layout is given, the objective function in \Cref{eq:new_objective} only takes into account the material cost. For the same reason, the line sizing constraint can be simplified to \Cref{eq:new_current}. \Cref{eq:new_tree} ensures that exactly one line type is chosen for each connection. The constraint for the voltage drop is reformulated in \Cref{eq:new_voltage}. The equation demands that the sum of all voltage drops in any path $p \in P$ from the source to a leaf must stay below the threshold $Q = U - U_{\mathrm{crit}}$. $P$ is the set of all paths from the source to a leaf in $\Gamma$. This updated formulation of voltage constraint is efficient, because all paths $P$ are given (due to the given layout), so the number of constraints resulting from \Cref{eq:new_voltage} remains low. 

\subsubsection{Solution Algorithm: Pairwise Edge Capacity Adjustment.}\label{sec:peca}

Since the solution of the previous binary integer program using commercially available solvers proves impractical for larger problem instances (as we show later in \cref{sec:cap_opt_eval}), we solve the capacity optimization problem via the following heuristic called \emph{pairwise edge capacity adjustment}~(PECA) heuristic. The PECA heuristic is based on two of our previously derived properties: first, \Cref{pro:decrease} states that the farther away from the source, the smaller the capacities become. Second, the heuristic utilizes \Cref{pro:ratio}, which defines the optimal ratio between two line capacities in the same path $p$ based on the flows, \ie $\frac{a_{ij}}{a_{mn}} = \sqrt{\tfrac{F_{ij}}{F_{mn}}}$. The PECA heuristic adjusts the capacities of two edges simultaneously in order to bring the ratio of these capacities as closely as possible to $\sqrt{\tfrac{F_{ij}}{F_{mn}}}$. Thereby, the heuristic increases the capacities of edges closer to the source and decreases the capacities of edges closer to the leaves.

The pseudocode is provided in \Cref{alg:PECA}. In \cref{line:for_p}, the heuristic iterates separately over each path $p \in P$ and then determines the capacities for that path as follows. In \cref{line:start_capacity_1,line:start_capacity_2}, the heuristic sets the capacities of all edges in $p$ to an initial value. This initial value is the minimum capacity such that the constraints for both line sizing and voltage drops are fulfilled. In \cref{line:first_half_1}, the heuristic loops over the depths $d$ of the edges in $p$. In \cref{line:first_half_2}, it selects an edge $(i,j)$ in the first half of $p$, that is, $d(j) \leq \tfrac{d(w)}{2}$ where $d(j)$ is the depth of vertex $j$. The corresponding edge $(m,n)$ further downstream in $p$ is determined in \cref{line:second_half}. This edge $(m,n)$ is as far away from the leaf $w$ as $(i,j)$ is from the source $0$, \ie $d(m) = d(w) - d(j)$. In \cref{line:peca_opt}, the heuristic determines the values for the capacities of the two edges $a_{ij}^{r^*}$ and $a_{mn}^{s^*}$ that minimize $\abs{\frac{a_{ij}^r}{a_{mn}^s} - \sqrt{\tfrac{F_{ij}}{F_{mn}}}}$. While doing so, the capacity $a_{ij}^{r}$ of the edge closer to the source is larger than or equal to its initial value $a_{ij}$, while the capacity $a_{mn}^{s}$ of the edge closer to the leaf is smaller than or equal to its initial value $a_{mn}$. Also, the constraints for both line sizing and voltage drops are fulfilled. This optimization problem can be solved via a strategic search by increasing $a_{ij}^{r}$ and/or decreasing $a_{mn}^{s}$ until $\abs{\frac{a_{ij}^r}{a_{mn}^s} - \sqrt{\tfrac{F_{ij}}{F_{mn}}}}$ does not get any smaller. In \cref{line:peca_set}, the optimized capacities are stored as candidate solutions. These values are compared with values from earlier iterations because an edge can be part of more than one path. In order to fulfill all constraints for all paths, we choose the maximum value for edges that have been optimized earlier in \cref{line:overwrite_2}. After that, the capacities of all edges have been optimized. The heuristic now conducts a post-hoc capacity adjustment in \crefrange{line:post_1}{line:post_end}. The reason is the following: some capacities might be larger than needed (\ie the previous minimization problem has introduced slack capacities), which can be reduced further in \cref{line:start_capacity_2,line:overwrite_2}.

	\begin{algorithm} 
	\SingleSpacedXI
		\caption{Pairwise edge capacity adjustment~(PECA)} \label{alg:PECA}
		\scriptsize
		\textbf{Input:} Network layout $\{x_{ij}\}$
		\begin{algorithmic}[1] 
			\For{each path $p$ connecting a leaf $w$ to the source $0$} \label{line:for_p}
				\For{each edge $(i,j) \in p$}			\label{line:start_capacity_1}
					\State $a_{ij}\gets \underset{k}{\min} \; \left\{ a_{ij}^{k} \;\middle|\; c_{ij}^k \geq F_{ij} \; \text{and} \; a_{ij}^k \geq \frac{1}{Q} \sum_{(i,j) \in p} l_{ij} F_{ij}  \right\} \quad$  
					\Comment{Set initial value for capacities}   \label{line:start_capacity_2}
				\EndFor
				\For {$d \in \{ 1, \ldots, \floor{d(w)/2} \}$}				\label{line:first_half_1}
					\State Select edge $(i,j) \in p$ with depth $d(j) = d$	\label{line:first_half_2} \Comment{Select edge in first half of $p$}
					\State Select edge $(m,n) \in p$ with $d(w)-d(m) = d(j)$			\Comment{Select corresponding edge in second half of $p$}	\label{line:second_half}
					\State \textbf{Compute} $r,s \gets \underset{r,s \in \{ 1, \ldots, \abs{A} \}}{\arg\min} \; \abs{\frac{a_{ij}^r}{a_{mn}^s} - \sqrt{\tfrac{F_{ij}}{F_{mn}}}}$ \label{line:peca_opt}	\Comment{Optimize capacities} 
					\Statex \quad \quad \quad \quad \textbf{s.\,t.} $a_{ij}^r \geq a_{ij}$, $a_{mn}^s \leq a_{mn}$,				\Comment{$a_{ij}$ can only be increased, $a_{mn}$ can only be decreased}
					\Statex \quad \quad \quad \quad \quad \quad $a_{mn}^s \geq F_{mn}$, $\sum\limits_{(v, w) \in p} l_{vw} \frac{F_{vw}}{a_{vw}} \leq Q$   \Comment{Line sizing and voltage drops must be fulfilled}
					\State $a_{ij} \gets a_{ij}^{r^*}, a_{mn} \gets a_{mn}^{s^*}$			\label{line:peca_set}		\Comment{Set both capacities to optimized values}
				\EndFor		
				\For {each edge $(i,j) \in p$}		\label{line:overwrite_1}
					\State $\widetilde{a}_{ij} \gets \max \{a_{ij}, \widetilde{a}_{ij}\}$ 	 	\Comment{Overwrite previous values, if necessary}		\label{line:overwrite_2}
				\EndFor
			\EndFor	
			\While {true}	\quad \label{line:post_1} \Comment{Post-hoc capacity adjustment} 
				\For {each edge $(i,j) \in \{\widetilde{E} \mid x_{ij} = 1 \}$}
					\State $k \gets \{ \kappa \mid \widetilde{a}_{ij} = a_{ij}^{\kappa} \}$ \label{line:lookup}	\Comment{Index lookup}
					\If{$k \neq 1$}
						\State $\widetilde{a}_{ij} \gets a_{ij}^{k-1}$ \label{line:peca_decrease}		\Comment{Decrease capacity}
						\If {$a_{ij}^{k-1} \geq F_{ij} \quad \textbf{and} \; \sum\limits_{(v, w) \in p} l_{vw} \cfrac{F_{vw}}{\widetilde{a}_{vw}} \leq Q \; \text{for all} \; \{ p \mid (i,j) \in p \}$} \label{line:peca_constraints_cleanup} 	\Comment{Check constraints}
							\State \textbf{continue}
						\Else {} $\widetilde{a}_{ij} \gets a_{ij}^{k}$		\label{line:peca_reset}	\Comment{Reset capacity, if constraints are violated}
						\EndIf
					\EndIf
				\EndFor
				\State \textbf{break}
			\EndWhile \label{line:post_end}
			\State \textbf{return} $\{\widetilde{a}_{ij}\}$
		\end{algorithmic}
	\end{algorithm}

The runtime of the PECA heuristic depends on the network layout. In a star network, every vertex is directly connected to the source. In this case, the PECA heuristic has a runtime of $\Theta(N)$ and returns the optimal solution to the capacity optimization problem. For other layouts, runtimes are higher as the number of leaves and the depth of the paths increase with a growing~$N$ \citep[\cf][]{Steele.1987}. 

\NEW{In Section~\ref{sec:proofs_PECA} of the supplements, we furthermore show that the error in terms of solution quality of the PECA heuristic is bounded and that the bound is tight in networks with similar line lengths.}


\subsection{Performance Evaluation} 
\label{sec:comp}

To demonstrate the effectiveness of our solution approach, we compared its performance in terms of solution quality and runtime against several benchmarks. These benchmarks address both the individual solution steps of optimizing layouts (\cref{sec:perf_eval_layout}) and line capacities (\cref{sec:cap_opt_eval}), as well as the solution of the full \pns (\cref{sec:perf_eval_complete_problem}). We summarize the results in the following; the numerical results are given in detailed tables in \cref{sec:evaluation_methods} of the supplements. 

\subsubsection{Layout Optimization.} \label{sec:perf_eval_layout}

For layout optimization, we compare the presented Tabu Search metaheuristic against two other heuristics, namely a \emph{variable neighborhood search~(VNS)} metaheuristic and an \emph{increased network branching~(INB)} heuristic. 

In general, neighborhood search methods explore the local neighborhood of the current solution to find a solution with a better objective value. Variable neighborhood search heuristics vary these local neighborhoods in order to find a better solution, as opposed to local search methods that do not use several neighborhoods within one method \citep[\cf][]{Hansen.2019}. In our case, the distance between neighboring solutions is defined by the difference in edges between two layouts. The principle of the INB heuristic is derived from network reinforcement practice: In practice, planners connect subparts of a network directly to the transformer in order to reduce voltage drops and peak flows. Thereby, the resulting reinforced network exhibits an increased branching in comparison to the original network layout. Details on both the VNS heuristic and INB heuristic are provided in \cref{sec:VNS} and \cref{sec:INB} of the supplements.


Our performance analysis shows that the Tabu Search metaheuristic is superior to the VNS and INB heuristics. The Tabu Search yields costs that are between 2\,\% and 4\,\% lower than costs obtained by the VNS. The cost advantage is consistent across network instances of different size. In comparison to the INB heuristic, the advantage of the Tabu Search is even larger, yielding consistent savings of between 8\,\% and 10\,\%. In summary, the results underline the effectiveness of the Tabu Search heuristic. Moreover, the high cost of the INB heuristic also implies that current practice in network planning is not optimal.

\subsubsection{Capacity Optimization.} \label{sec:cap_opt_eval}	

For network capacity optimization, we compare our PECA~algorithm against exact solutions from a \emph{MIP solver} and two greedy heuristics informed by network planning practice, namely \emph{greedy capacity reinforcement} and \emph{greedy capacity reduction}. 

The MIP solver makes use of the problem stated in \Crefrange{eq:new_objective}{eq:new_voltage}. The problem represents a binary integer program which is then solved using the Gurobi Optimizer~7.5.2. 
The greedy capacity reinforcement heuristic starts with minimal edge capacities and optimizes the capacities by steadily increasing them until all capacity constraints are fulfilled. It starts with minimal edge capacities, successively identifies the edge $(i, j)$ with the highest voltage drop (this corresponds to the weak spot of the network) and then increases the capacity of this edge. 
The greedy capacity reduction heuristic is the counterpart to the greedy capacity reinforcement heuristic. It starts with the largest edge capacities and proceeds in the opposite direction: It identifies the edge $(i, j)$ with the lowest voltage drop and reduces its capacity in order to save material cost. Details on both heuristics are provided in \cref{sec:greedy_reinforcement} and \cref{sec:greedy_reduction} of the supplements.


In terms of solution quality, we find that the PECA algorithm consistently outperforms both the greedy capacity reinforcement heuristic and the greedy capacity reduction heuristic, leading to average cost savings of between 2\,\% and 12\,\%. In comparison to the MIP solver, the PECA algorithm is largely on par in terms of network cost, with an average optimality gap of less than 1\,\% for small networks and less than 2\,\% for large networks. However, in terms of runtime, the PECA algorithm outperforms the exact solver considerably for all network sizes: The PECA algorithm yields average runtime advantages against the MIP solver by a factor of two to six (depending on network size and topology). 

Capacity optimization is responsible for a crucial part of the overall runtime for solving \pns. This is due to the fact that, during layout optimization, the improvement heuristic triggers a capacity optimization for each candidate layout. Hence, runtime savings in the capacity optimization will lead to large improvements in overall solution speed. Because of this, we use the PECA algorithm for all subsequent analyses as it has a considerably better runtime than the alternatives. 


\subsubsection{Full \pns.} \label{sec:perf_eval_complete_problem} 	

In order to evaluate the effectiveness of the above solution approach in solving the full \pns, we use \NEW{five} additional benchmarks:
\begin{itemize}
    \item First, we linearize the \pns (as described in \Cref{sec:lin}) and solve it using the \emph{Gurobi MIP solver}. 
    \item{\NEW{Second, we provide an alternative model formulation for the \pns (``\pns with undiscounted flows''). Instead of discounting the flows with the coincidence factor, this model is based on \emph{undiscounted} flows. The model, including linearizations, is provided in \Cref{sec:commodities_model} of the supplements. Solutions are computed using the \emph{Gurobi MIP solver.}}} 
    \item \NEW{Third,} we perform a \emph{complete enumeration} of layouts and capacities. For both approaches, in case an optimal solution could not be found within the given time limit, we evaluated the best solution obtained. 
    \item \NEW{Fourth and fifth,} we compute an \emph{upper} and \emph{lower bound} for the exact solution using simplified problem instances. The bounds are determined by constantly over- or underestimating the demand using a uniform coincidence factor for all edges in the network. This results in a simplified version of the \pns that was previously formalized as a distribution network reconfiguration problem \citep[\eg][]{Avella.2005}, which we then solve using the Gurobi Optimizer MIP solver. For the upper bound, we choose a coincidence factor of $\gamma \equiv 1$. This corresponds to a scenario where the peak demand is fully coinciding. For the lower bound, we apply a coincidence factor of $\gamma \equiv \gamma(N-1)$ as an overall discount factor to all loads. This corresponds to the maximum achievable discount for a network of size $N$.
\end{itemize}
We find that, due to computational complexity, exact solutions cannot be obtained for all network instances, neither through linearization nor enumeration. \NEW{Only for very small network instances ($N=5$) is the MIP solver able to find optimal solutions (for both the \pns and the \pns with undiscounted flows). For network instances with 20 vertices or more, the MIP solver is not able to determine any feasible solution for either model. The solution approach based on the Tabu Search, on the contrary,} remains within a distance of 2\,\% to 7\,\% of the lower bound. Moreover, our solution approach reaches costs that remain between 7\,\% and 48\,\% of the upper bound. In summary, these results underline the effectiveness of our solution approach for solving the complete \pns. 

\section{Impact of Demand Coincidence on the Cost of Electricity Distribution Networks} 
\label{sec:impact}

In this section, we evaluate the impact of coinciding demand on the cost of electricity distribution networks. In \cref{sec:case}, we conduct an evaluation using real-world network instances from a Swiss electricity company in order to determine the impact on actual investment cost. To isolate the effects of different network characteristics, we conduct an additional evaluation in \cref{sec:num} based on synthetic network instances, where we generate networks with different characteristics. 


\subsection{Evaluation Using Real-World Networks}
\label{sec:case}

\subsubsection{Network Statistics.}

Our real-world experiments are based on a sample of 74 low voltage distribution networks from a Swiss distribution network operator. Each network entails one transformer. The number of loads per network in the sample ranges from 12 to 68, with an average of 36.7~loads and a median of 32~loads. The loads correspond to the vertices in our model. The costs for the networks range from CHF~33,500 to CHF~1.7~million. The average network cost is CHF~251,400. The networks sum up to a combined value of CHF~18.6~million. This cost only includes material and construction and excludes planning and overhead costs. On average, each network covers an area of \SI{35.1}{\hectare}, \ie \SI{0.351}{\square\kilo\metre}, with a median size of \SI{20.6}{\hectare} per network.

\subsubsection{Setup.} \MOVED{For our experiment, we use network and demand data from our partnering utility company.} \NEW{The objective is to identify among the edges of the complete multigraph formed by interconnecting all locations with all available cable types (i.e. $(i,j)^k \in V \times V \times K$) the cost-optimal configuration.}\footnote{\REVTWO{This setup corresponds to a greenfield approach, which has the advantage that one determines the effect of increased load coincidence independent of the slack capacities of pre-existing networks. The brownfield problem can be studied as well, by setting $c_{\mathrm{c}} = c_{\mathrm{m}} = 0$ for preexisting lines and cable types. In this case, the capacity optimization subproblem should be solved using mixed-integer programming instead of the PECA algorithm (as described in \Cref{sec:add_info_p2} of the supplements.)}} \MOVED{Thereby, we compare different scenarios of load coincidence.

For the \emph{objective function}, we use the cost constants $c_{\mathrm{c}} = 34.62 \, \tfrac{\mathrm{CHF}}{\mathrm{m}}$ and $c_{\mathrm{m}} = 0.1882 \, \tfrac{\mathrm{CHF}}{\mathrm{m \; mm^2}}$. These follow the real-world cost composition and are based on our partnering company’s engineering department’s cost sheets as well as discussions with network design experts.}


The \emph{network data} (\ie longitude and latitude of nodes, consumer loads) has been extracted from our partner company's geographic information system. All distances between locations are computed using the Euclidean distance (L2 norm). Some of the locations in the original data set belong to components with zero energy demand (\eg fuse boxes), which are included with $D_i = 0$. \MOVED{For all consumer locations, we use the industry standard of $D^{\mathrm{peak}} = \SI{21}{kW}$ for the peak demand per load \citep{kaufmann1995planung, schulz2010elektrische}.}  An example network layout is shown in \Cref{fig:data_preparation}. 

\begin{figure}[ht]
\SingleSpacedXI
\centering
\includegraphics[width = .6\textwidth]{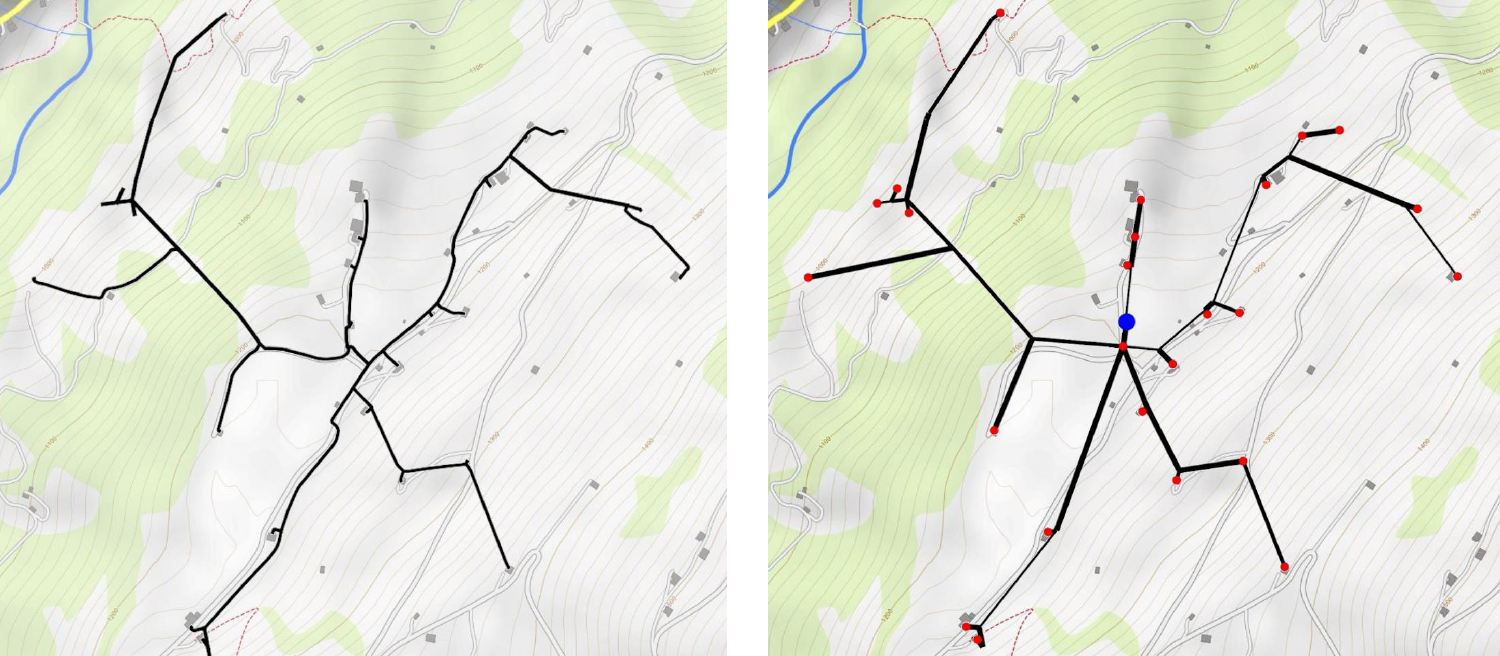}
\caption{\raggedright Example real-world network with buildings, roads, and landscape (left) and pre-processed network layout (right) with transformer (large dot) and consumers (small dots).}
\label{fig:data_preparation}
\end{figure}

\MOVED{All \emph{cable characteristics}~(capacities, electrical resistance, and costs) have been set to conventional values from practice in agreement with practitioners from our partner company. Specifically, for the line capacities, we use values corresponding to commercially available copper cables, identical to the ones used in the original networks. The capacity of these cables is given as a cross section in \SI[parse-numbers = false]{}{\square\milli\metre}. Capacities $a_{ij}^k \in A$ are chosen from the set $A = \{ \SI{50}{\square\milli\metre}, \SI{70}{\square\milli\metre}, \SI{95}{\square\milli\metre}, \SI{120}{\square\milli\metre}, \SI{150}{\square\milli\metre}, \SI{185}{\square\milli\metre}, \SI{240}{\square\milli\metre}, \SI{400}{\square\milli\metre}, \allowbreak \SI{800}{\square\milli\metre} \}$. \allowbreak The maximum number of allowed currents per line type are given in \SI[parse-numbers = false]{}{\ampere}. The current carrying capacities $c_{ij}^k \in C$ can be chosen from $C = \{ \SI{185}{\ampere}, \SI{228}{\ampere}, \SI{274}{\ampere}, \SI{313}{\ampere}, \SI{352}{\ampere}, \SI{398}{\ampere}, \SI{464}{\ampere}, \allowbreak \SI{510}{\ampere}, \SI{671}{\ampere} \}$. The value for the resistivity of the network lines, $\rho = 0.0181 \tfrac{\Omega \mathrm{mm}^2}{\mathrm{m}}$ has been extracted from the data sheet of the electrical cables used.} 

We model the \emph{coincidence factor} via $\gamma(N) = \gamma_{\lim} + (1-\gamma_{\lim}) \, N^{-1/2}$ \citep{Rusck.1956, Dickert.2010}. In the following, we vary the parameter $\gamma_{\lim}$ to create three scenarios of different degrees of demand coincidence. (For each scenario, we can use the corresponding function $\gamma_{\gamma_{\lim}}(N)$ to calculate coincidence factors for arbitrary sets of consumers. A larger (smaller) $\gamma_{\lim}$ also implies a larger (smaller) coincidence factor.) 
The first scenario uses the parameter $\gamma_{\lim} = 0.1$, 
which follows current practice in network planning and serves as the baseline scenario for our numerical experiments \citep[see, \eg][]{Dickert.2010}.
The second scenario considers a future setting with a larger demand coincidence. This should reflect future developments, \eg due to electric vehicles that are charged concurrently \citep[\eg][]{Verzijlbergh.2011}. For this, we set $\gamma_{\lim} = 0.7$. 
The third scenario represents a worst case where all demand peaks take place at the same time (\ie $\gamma_{\lim} = 1.0$). This results in a constant coincidence factor of $\gamma(|\Gamma_j|) \equiv 1.0$. Hence, this scenario estimates the worst-case effect of demand coincidence on network cost. Research has found that this can be a realistic scenario for uncontrolled charging of electric vehicles: \Citet{Gaul.2017}, for instance, derive a coincidence factor of $1.0$ after analyzing 450,000~charging sessions at public charging stations in Germany. \NEW{Since residential charging stations for electric vehicles need power of up to \SI{20}{kW} \citep{rachid2023, deilami2020insight} --- which is roughly equivalent in size to the industry standard of $D^{\mathrm{peak}} = \SI{21}{kW}$ peak demand per consumer location \citep{kaufmann1995planung, schulz2010elektrische}, \REVTWO{this simultaneous charging can create new demand peaks} and thus appears as a reasonable worst-case scenario.}

For ease of understanding, we illustrate the objective function and main constraints under the given parameter settings in Section~\ref{sec:add_info_case} of the supplements. Furthermore, all parameters used as part of the heuristics are identical to the previous computational experiments.

\subsubsection{Results.}

\Cref{tab:case_study} compares the cost and the length of the networks for the described scenarios of varying demand coincidence. The networks are grouped by the number of loads $N$ contained in each network, ranging from $N \in [10,19]$ to $N \in [60,69]$. The number of networks in each group is reported alongside our results. 
All costs are given in Swiss Francs~(CHF). We report average costs for each of the six groups.

\begin{table}[htbp]
  \centering
  \caption{Results for real-world networks comparing different scenarios of demand coincidence.}
  \resizebox{\columnwidth}{!}{
    \begin{tabular}{llrrrrrr}
    \toprule
          &       & \multicolumn{3}{l}{\textbf{Panel A: Network cost (in CHF)}} &       &       &  \\
    \midrule
    \textbf{Scenario} & \textit{\textbf{N $\in$}} & \textbf{[10, 19]} & \textbf{[20, 29]} & \textbf{[30, 39]} & \textbf{[40, 49]} & \textbf{[50, 59]} & \textbf{[60, 69]} \\
    \midrule
    1: Baseline ($\gamma_{\lim} = 0.1$) &       & 58,659 & 102,732 & 68,831 & 140,317 & 82,988 & 313,230 \\
    \midrule
    2: Increased demand coincidence ($\gamma_{\lim} = 0.7$) &       & 73,431 & 170,058 & 144,000 & 258,160 & 200,052 & 440,991 \\
    \midrule
    3: Worst case ($\gamma_{\lim} = 1.0$) &       & 82,350 & 177,729 & 164,927 & 265,218 & 215,318 & 493,987 \\
    \midrule
    \midrule
    \textit{Number of networks in sample group} &       & \textit{14} & \textit{19} & \textit{9} & \textit{9} & \textit{14} & \textit{9} \\
    \midrule
        \multicolumn{8}{l}{Stated: average per sample group }\\
          &       &       &       &       &       &       &  \\
          \\
          \toprule
          &       & \multicolumn{4}{l}{\textbf{Panel B: Length in km (no. of branches)}} &       &  \\
\toprule 
\textbf{Scenario} & \textit{\textbf{N $\in$}} & \textbf{[10, 19]} & \textbf{[20, 29]} & \textbf{[30, 39]} & \textbf{[40, 49]} & \textbf{[50, 59]} & \textbf{[60, 69]} \\
    \midrule
    1: Baseline ($\gamma_{\lim} = 0.1$) &       & 1.290 (2.1) & 2.017 (1.9) & 1.409 (1.3) & 2.806 (2.3) & 1.750 (1.9) & 6.373 (8.2) \\
    \midrule
    2: Increased demand coincidence ($\gamma_{\lim} = 0.7$) &       & 1.541 (3.1) & 3.467 (6.0) & 3.020 (6.9) & 5.335 (9.3) & 4.484 (10.6) & 8.490 (13.9) \\
    \midrule
    3: Worst case ($\gamma_{\lim} = 1.0$) &       & 1.713 (3.6) & 3.57 (6.6) & 3.388 (7.7) & 5.441 (9.7) & 4.776 (12.4) & 9.150 (15.4) \\
    \midrule
    \midrule
    \textit{Number of networks in sample group} &       & \textit{14} & \textit{19} & \textit{9} & \textit{9} & \textit{14} & \textit{9} \\
    \bottomrule
    \multicolumn{8}{l}{Stated: average per sample group }\\
    \end{tabular}}%
  \label{tab:case_study}%
\end{table}%

In general, the cost per network increases with network size. However, in rare situations, networks with fewer consumers can result in a larger cost than networks with more consumers, which is due to geographic circumstances. 
Demand coincidence has a profound impact on the cost of a network and its layout. For example, for the largest instances with $N \in [60, 69]$, the cost difference between the worst case scenario and the baseline scenario is CHF~0.2~million per network. Here, network lines are up to 3~km longer.
In general, a larger demand coincidence consistently leads to more expensive networks. On average, the cost difference between the baseline case  ($\gamma_{\lim} = 0.1$) and the worst case ($\gamma_{\lim} = 1.0$) is 84~\%. In the majority of cases, cost differences increase with network size, ranging from 40\,\% in the group of smallest networks (10 and 19 loads) to up to 159\,\% in the group of networks with 50 to 59 loads. 

Regarding the effect of demand coincidence on network layouts, we find that the increase in network cost can almost entirely be attributed to an increase in the length of the network. This means the larger the demand coincidence, the longer the networks (not necessarily the higher the demand coincidence, the larger the line capacities). For instance, for networks with $N \in [10, 19]$, $N \in [20, 29]$, and $N \in [30, 39]$, the average increase in cost between the baseline scenario  ($\gamma_{\lim} = 0.1$) and the worst case scenario ($\gamma_{\lim} = 1.0$) is 40\,\%, 73\,\%, and 140\,\% respectively. The average increase in length is 33\,\%, 77\,\%, and 140\,\% respectively, while the average capacities remain largely unchanged. 

In summary, we find that the effect of demand coincidence on network cost is considerable. For network instances with more than $30$ loads, we find that fully coinciding demand can easily double the network cost.  

\subsection{Evaluation Using Synthetic Networks}
\label{sec:num}

We now study how the impact of demand coincidence changes across networks with different characteristics. For this, we perform an additional evaluation based on synthetic network instances, where we specifically generate networks with certain characteristics.

\subsubsection{Setup.} 

The experiments are conducted for synthetic network instances of different size $ N \in \{20, 40 \ldots, 100\}$. For each $N$, we generate 20 instances and later report average costs and line lengths. \NEW{The $x$- and $y$-locations of the vertices $(s_x, s_y)$ are sampled from a discrete uniform distribution without replacement. This ensures that consumers have different, non-overlapping locations as well as a certain minimum distance between them. We vary the distance between customers, so that we distinguish three cases of sparsely/densely populated neighborhoods: (1)~a case with a low consumer density of $0.1$ consumers per area unit, (2)~a medium case with $1$ consumer per area unit, and (3)~a high density case with $10$ consumers per area unit.} We set $U_{\mathrm{crit}}$ to a value corresponding to reality (\eg as used by our partner company). This and all other parameters are listed in \Cref{tab:parameters}. For simplicity, we omit units. (The appropriate unit conversions are in \Cref{sec:add_info_case} of the supplements.) 


Importantly, we again study different scenarios of demand coincidence. For this, we vary $\gamma_{\lim} \in \{0.1, 0.5, 1.0\}$, representing scenarios of low, medium, and high demand coincidence. All details on the implementation are in \Cref{sec:add_info_comp} of the supplements.


\begin{table}[ht]
\caption{Parameters for computational experiments.} 
\label{tab:parameters}
{\resizebox{\columnwidth}{!}{
\begin{tabular}{cccccc}
\toprule
\begin{tabular}[c]{@{}c@{}}Network \\ size \end{tabular} & \begin{tabular}[c]{@{}c@{}} Edge \\ capacities   \end{tabular}  & \begin{tabular}[c]{@{}c@{}} Voltage drop \\ threshold  \end{tabular} & \begin{tabular}[c]{@{}c@{}} Peak \\ demands \end{tabular} & \begin{tabular}[c]{@{}c@{}} Coincidence \\ factor \end{tabular} & \begin{tabular}[c]{@{}c@{}}  Costs \end{tabular}  \\ 
\midrule
$ N \in \{20, 40, \ldots, 100\}$ & \begin{tabular}[t]{@{}c@{}}$a_{ij}^k \sim \{0.1, 0.2, \ldots, 1.0\}$ \\ $c_{ij}^k \sim \{0.1, 0.2, \ldots, 1.0\}$ \end{tabular} & $U_{\mathrm{crit}} = U - 1.0$ & $D^{\mathrm{peak}}=0.01$ & $\gamma(|\Gamma_j|) = \gamma_{\lim} + (1 - \gamma_{\lim}) \, |\Gamma_j|^{1/2}$ & $c_c = c_m = 1$     \\ \bottomrule
\end{tabular}}
}
{}
\end{table}

\subsubsection{Results.}

The numerical results are in \Cref{tab:nw_cost}. As in the real-world evaluation, the cost of networks increases in the number of vertices for all scenarios. Here, large networks ($N = 100$) are 6 to 16 times more expensive than small networks ($N = 20$). 
For demand coincidence, we observe increasing cost for a larger $\gamma_{\lim}$ across all network sizes. For small networks, the cost difference between the baseline scenario with $\gamma_{\lim} = 0.1$ and the worst case with $\gamma_{\lim} = 1.0$ is up to 22$\,\%$. For the largest networks, this difference reaches 81\,\%, \ie an almost four-fold increase. 


\begin{table}[htb]
\caption{Results from synthetic networks across different scenarios of demand coincidence.} 
\centering
\footnotesize
\label{tab:nw_cost}
{\resizebox{0.65 \columnwidth}{!}{ 
\setlength{\aboverulesep}{0pt}
\setlength{\belowrulesep}{1pt}
\renewcommand{\arraystretch}{1.0}
    \begin{tabular}{lllrrrrr}
    \toprule
    \textbf{Density} & \textbf{Scenario ($\gamma_{\mathrm{lim}}$)} & \textit{\textbf{N\,$=$}} & \textbf{20} & \textbf{40} & \textbf{60} & \textbf{80} & \textbf{100} \\
    \midrule
    \multirow{3}{*}{Low} & 0.1   &       & 56.9  & 147.5 & 260.2 & 438.6 & 602.8 \\
\cmidrule{2-8}          & 0.5   &       & 63.0 & 184.3 & 360.9 & 635.7 & 849.1 \\
\cmidrule{2-8}          & 1.0   &       & 69.6 & 228.5 & 419.4 & 793.6 & 1,092.1 \\
    \midrule
    \multirow{3}{*}{Medium} & 0.1   &       & 15.8  & 34.2  & 52.3  & 76.2  & 103.6 \\
\cmidrule{2-8}          & 0.5   &       & 16.5 & 37.9 & 58.9 & 92.0 & 131.0 \\
\cmidrule{2-8}          & 1.0   &       & 17.3 & 42.3 & 67.1 & 108.6 & 160.8 \\
    \midrule
    \multirow{3}{*}{High} & 0.1   &       & 4.8   & 9.9   & 15.3  & 21.4  & 26.6 \\
\cmidrule{2-8}          & 0.5   &       & 4.8 & 10.3 & 16.2 & 23.3 & 29.2 \\
\cmidrule{2-8}          & 1.0   &       & 4.9 & 10.8 & 17.4 & 25.2 & 31.7 \\
    \bottomrule
    \multicolumn{8}{l}{\footnotesize Stated: cost averaged over 20 random instances per combination of density and scenario}\\
    \end{tabular}}
}
{}
\end{table}


We now analyze how the impact of demand coincidence varies across networks of different densities. Generally, a higher density among loads results in less expensive networks. This behavior is to be expected---in more dense networks, lines are shorter on average, because distances are smaller; additionally, smaller line capacities suffice, because less voltage drops accumulate.
For demand coincidence, we observe that, in networks with a low density, coincidence factors play an important role. For example, for large networks ($N = 100$), the relative cost differences between the baseline scenario and the worst case scenario are around 81$\,\%$ for the low density, 55$\,\%$ for the medium density, and 19$\,\%$ for the high density case. Consistent patterns are also found for smaller network sizes, where the cost increase due to a larger demand coincidence is higher for low than for high density networks. Translated to real-world settings, this means that the effect of demand coincidence is more pronounced in sparsely populated, rural areas. 

\NEW{We further analyze the sensitivity of the resulting networks with regard to the assumed cost parameters. More precisely, we examine how the resulting networks depend on the ratio of assumed material cost $c_m$ and construction cost $c_c$. For this purpose, we keep $c_c = 1$ and vary $c_m \in \{0.1,0.5,\ldots, 100\}$. We then analyze how increasing demand coincidence affects the network characteristics under the varying cost parameters. From the resulting networks, \cref{tab:sens_cost} displays the total number of line lengths and the average line capacities. By comparing the values within individual columns, we can understand how the resulting network structures are affected by increasing demand coincidence. By comparing across columns, we can then study how these effects change under different cost parameters.}

\begin{table}[htbp]
  \centering
  \caption{\NEW{Network line capacities (Panel A) and line lengths (Panel B) under varying cost parameters.}} 
  \label{tab:sens_cost}
  \resizebox{.7\columnwidth}{!}{
    \begin{tabular}{llrrrrrrrr}
    \toprule
          &       & \multicolumn{8}{l}{\textbf{Panel A: Average line capacities}}  \\
    \midrule
    \textbf{Scenario ($\gamma_{\lim}$}) & \textit{\textbf{$c_m =$}} & \textbf{0.1} & \textbf{0.5} & \textbf{1} & \textbf{5} & \textbf{10} & \textbf{20} & \textbf{50} & \textbf{100} \\
    \midrule
    $\gamma_{\lim} = 0.1$ &       & 0.139 & 0.126 & 0.112 & 0.096 & 0.095 & 0.094 & 0.093 & 0.092\\
    \midrule
    $\gamma_{\lim} = 0.5$ &       & 0.218 & 0.165 & 0.146 & 0.122 & 0.112 & 0.108 & 0.103 & 0.100\\
    \midrule
    $\gamma_{\lim} = 1$ &       & 0.274 & 0.231 & 0.215 & 0.151 & 0.132 & 0.120 & 0.124& 0.109\\
    \midrule
    \midrule
        \multicolumn{10}{l}{Stated: averages over 15 random instances per combination of $c_m$ and scenario}\\
          &       &       &       &       &       &       &  &&\\
          \\
          \toprule
          &       & \multicolumn{8}{l}{\textbf{Panel B: Total network line length}}\\
\toprule 
    \textbf{Scenario ($\gamma_{\lim}$}) & \textit{\textbf{$c_m =$}} & \textbf{0.1} & \textbf{0.5} & \textbf{1} & \textbf{5} & \textbf{10} & \textbf{20} & \textbf{50} & \textbf{100} \\
    \midrule
    $\gamma_{\lim} = 0.1$ &       & 16.272 & 16.340 & 16.448 & 17.002 & 17.377 & 17.784 & 17.948 & 18.198\\
    \midrule
    $\gamma_{\lim} = 0.5$&       & 16.289 & 16.554 & 16.767 & 17.935 & 18.767 & 20.033 & 20.970 & 21.169\\
    \midrule
    $\gamma_{\lim} = 1$ &       & 16.278 & 16.690 & 16.905 & 18.663 & 20.941 & 22.899 & 23.858 & 25.841\\
    \midrule
    \midrule
    \multicolumn{10}{l}{Stated: averages over 15 random instances per combination of $c_m$ and scenario}\\
    \end{tabular}}%
\end{table}%

\NEW{We observe that, under low values of the material cost parameter $c_m$, networks are expanded to cope with higher demand coincidence by adding larger line capacities (see Panel A) and not by increasing the network branching (which would result in larger line lengths in Panel B---however, these remain almost unchanged for $c_m \leq 1$). This appears natural, since using lines of a higher capacity is, ceteris paribus, the cheaper expansion measure. Under higher values of the material cost parameter ($c_m \geq 20$), the results are in opposite direction, \ie increasing the branching of the networks is more cost-effective. Finally, we emphasize that the cost ratios are varied considerably (within three orders of magnitude); yet, we observe gradual and monotonous behavior of the resulting network capacities and line lengths. This means that the resulting networks are rather insensitive to the cost parameters, \ie large changes in the parameters are required to affect the resulting network layouts and capacities. 
}

Finally, we look at the nature of the effect of more coincident demand on network cost. This is shown in \cref{fig:cost_v_cf} for networks with $N=60$ and of medium density. (The findings are similar for networks of a different size or density). The plots show that, on average, the cost of networks is characterized by a close-to-linear behavior. At the level of individual network instances, the increasing trend remains visible; however, cost increases are more heterogeneous: For some values of $\gamma_{\mathrm{lim}}$, a slight increase in demand coincidence can result in a large change in cost, whereas, for other $\gamma_{\mathrm{lim}}$, there is almost no effect. This can be explained by the fact that, in some cases, the increase in demand coincidence requires considerable changes in the network topology, whereas, in other cases, the network configuration is able to withstand it and only minor expansions of line capacities are required, thus explaining why individual curves are not straight lines but show some fluctuations. 

\begin{figure}[htbp]
    \centering
      \SingleSpacedXI
	\footnotesize 
    \includegraphics[width = 0.5\textwidth]{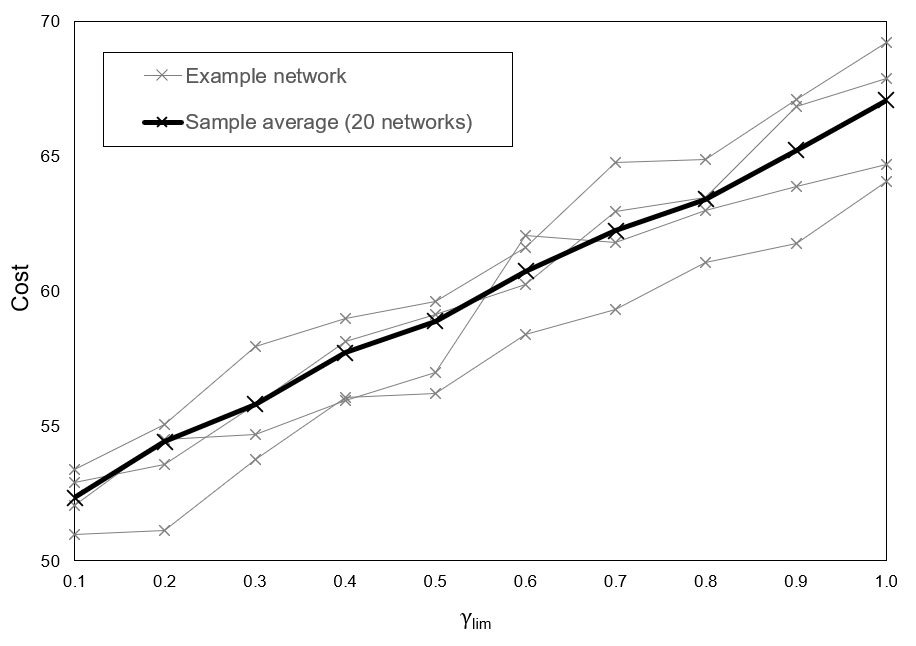}
    \caption{\raggedright Impact of demand coincidence on the cost of networks. Results averaged over 20 random network instances (for $N=60$ and a medium density). Out of the 20 instances, four random examples are shown in gray.}
    \label{fig:cost_v_cf}
\end{figure}


In summary, the above evaluation confirms the finding from the real-world networks: increasing the coincidence of electricity demand has a large effect on the cost of networks. Similarly, the effect is more pronounced for larger networks. We additionally observe that effects are more pronounced for networks with a low density of load as in sparsely populated neighborhoods. 


\section{Concluding Discussion} 
\label{sec:disc}

We structure our discussion along the main contributions of this work: In \cref{sec:disc_findings}, we discuss our general findings regarding the effect of increased demand coincidence on network investments. 
In \cref{sec:disc_findings_structures}, we continue by discussing our findings regarding the nuanced effects of increased coincidence on different network structures. We then elaborate on our methodological contributions in \cref{sec:disc_method_contribution} and how these can support network operators in practice, \NEW{before we conclude by discussing limitations of our work in \cref{sec:limitations}.} 


\subsection{Effect of Increased Demand Coincidence on Network Investments}
\label{sec:disc_findings}

Our main finding is that the increased coincidence of electricity demand has a strong impact on the necessary investments when designing electricity distribution networks. In the evaluation with real-world networks, our model suggests that the difference in required investments between low and high degrees of coincidence is, on average, 84\,\%. Furthermore, the strong impact of demand coincidence on network cost is also confirmed by the numerical evaluation using synthetic networks. 


\NEW{Related research has arrived at similar findings. Regarding electric vehicles, the impact on distribution network reinforcements is confirmed by several studies \citep[see, \eg][for overviews]{dharmakeerthi2011overview, ahmad2022optimal}. For instance, \citet{gupta2021spatial} show that, under an electric vehicle penetration rate of 33\,\%, about 10\,\% of load connection points, 50\,\% of transformer stations, and several hundreds of kilometers of line segments need reinforcements. Similarly,  \citet{fernandez2010assessment} find that the integration of electric vehicles requires additional investment cost in distribution networks of up to 20\,\%. 
Regarding other load types, \citet{Hoflich.2012} analyze the impact of the coordinated execution of controllable loads---such as energy storages---and find that by 2030 these can cause an increase of about 40\,\% in additional investments. Similarly, \citet{Gwisdorf.2010} show that a penetration of controllable loads to a share of more than 20\;\% of households causes a strong increase in network reinforcements. Analogously, regarding heat pumps, \citet{gupta2021spatial} show that, under a full penetration rate, about 33\;\% of distribution networks require reinforcement investments. Regarding photovoltaic systems, \citet{Gust.2014} and \citet{Gust.2016} find that photovoltaic penetration rates of about 30\,\% cause overloads exceeding network capacities.}


\NEW{In summary, related studies agree that increasing demand coincidence has a strong impact on network investments. Nevertheless, there are differences in the magnitude of the estimated impact, which may be caused by different factors. First, these may be due to different assumptions. For instance, we assume a worst-case scenario of full load coincidence, whereas other studies assume smaller penetration rates of new technologies that likely affect coincidence less strongly. For instance, \citet{gupta2021spatial} and \citet{ fernandez2010assessment} assume electric vehicle penetration rates of 33\,\%  and 60\,\% respectively, whereas our results correspond to a full penetration of 100\,\%. Second, differences may also be caused by the methodological approach. Most of the previous findings have been generated on a set of reference networks. These are created by analyzing key characteristics of real-world distribution networks and then identifying (or constructing) networks with representative characteristics for the set or subset \citep[cf.][]{miguez2002improved, mateo2018european, Domingo.2011, cohen2016effects}. Finally, the findings also depend on the experimental parameter settings. Even though the parameters in our study were chosen deliberately together with experts from our partnering utility company, these may not be representative of all real-world circumstances. Nevertheless, despite the differences in models, methods, and parameter assumptions, the studies agree on a qualitative level that demand coincidence, if not counteracted through appropriate measures, will make considerable investments in distribution networks necessary.}  

In the following, we briefly discuss the implications resulting from the above finding. \TEMPERED{Policy-makers may need to pay additional attention to}  the distribution stage in the energy supply chain \citep[cf.][]{Kleindorfer.2005} by designing policies and operational protocols that reduce demand coincidence, particularly as energy use is increasingly electrified. \NEW{In this regard, policy-makers and regulators may choose from a considerable set of tools \citep[see also the MIT Utility of the Future study by][]{MITUtilityOfTheFuture}: First, smart pricing schemes may be promoted that incentivize the temporal diversification of consumer peak loads. Such schemes include non-linear, capacity-based pricing approaches \citep[\eg][]{Valogianni.2020,gottwalt2015managing}, or pricing schemes that include a network fee based on available network capacity in the electricity bill, such as nodal or area pricing \citep[\eg][]{liu2018} and similar approaches that apply local and temporal differentiation \citep{MITUtilityOfTheFuture}. However, given the local proximity of consumers in distribution networks, the local differentiation must be highly granular---which may be challenging to achieve.
Alternatives beyond pricing schemes include new market designs and smart connection agreements: Local energy markets can, for instance, be designed so that localized load balancing within a specific area or community can be achieved \citep[\cf][]{STROHLE2016811}. Smart connection agreements include load monitoring and control approaches where the grid operator can have the ability to remotely control or manage selected loads during critical periods, such as the ability to reduce the charging rate of electric vehicles \citep{connectingthedots}. Similarly, consumers may be incentivized to participate in critical peak pricing or similar demand response programs that aim to reduce grid congestion \citep[\cf][]{bossmann2016model, DoE2006}. 
}

For managers and operators of electricity distribution networks, the findings \TEMPERED{indicate} to take into account demand coincidence in long-term network planning tasks. The results show that investment budgets \TEMPERED{may need to be} adapted depending on the expected degree of demand coincidence of future technologies. For instance, in the case of electric vehicles, several studies have projected highly coincident electricity consumption  \citep[\eg][]{Gaul.2017,Verzijlbergh.2011}. Our real-world study shows that such scenarios of high coincidence will likely translate into large increases in network cost. 

\subsection{Effect of Increased Demand Coincidence on Network Structures}
\label{sec:disc_findings_structures}


In addition to the overall impact on network investments, we find that a high coincidence factor has a stronger impact on network layouts than on capacities. In our real-world study, we find that the average line lengths within networks increase between the baseline and the worst case of demand coincidence by an average of 82\,\%---\ie they almost double, whereas line capacities remain largely the same. \NEW{Our numerical sensitivity analysis also underlines that moderate variations in the cost parameters, \eg as caused by different procurement costs, lead to consistent results.} For managers of electricity distribution networks, this implies that an increase in demand coincidence requires a substantial installation of new network lines. Unlike line capacity expansions for which cables are replaced using the pre-existing infrastructure in place (such as overhead poles or underground pipes), the installation of new lines requires larger construction measures (such as re-wiring and trenching work).


Finally, we find that the effect of demand coincidence differs considerably depending on network characteristics. Our numerical study shows that increases in demand coincidence affect networks with a low consumer density particularly strongly. In such networks, a large coincidence factor increases network cost by up to 81\,\%. For comparison, the increase amounts to only 19\,\% for high density networks. Additionally, networks with a larger number of consumers are also more strongly affected. For example, for a low consumer density, this yields cost increases of up to 81\,\% in comparison to only 22\,\% for small networks.


The latter finding shows that, when electricity consumption becomes more coincident, effects on network infrastructures vary in different geographic areas \citep{Geis.2017}. Since consumer density is lower in rural areas, these generally require larger investments. Villages and other rural areas are thus confronted with overproportional increases in network costs, because, in such areas, networks are typically rather large and exhibit low density. Distribution networks in cities have comparatively smaller cost increases due to demand coincidence because of a generally higher consumer density. 
\NEW{Related research has arrived at similar conclusions. \Citet{gupta2021spatial} show that the grid integration of electric vehicles, heat pumps, and photovoltaic systems requires more reinforcement investments in rural networks than in their urban counterparts. Analogously, \citet{MCKINNEY2023} deem rural grids less robust for the integration of electric vehicles.} For policy-makers, this means that technologically-induced changes in the coincidence factor \TEMPERED{may} require unequal infrastructure investments. Particularly, rural areas are likely to require a larger investment budget. For this reason, policy-makers \TEMPERED{may} need to consider whether distribution network operators in rural areas need to be regulated differently than their urban counterparts to avoid unfairness. For instance, this could be achieved by including a geographic coefficient in regulation schemes. 
\NEW{However, in many countries, rural areas are already today cross-subsidized to avoid excessive network charges \citep[][]{BNetzA,oas}. Covering the additional effects of more coincident demand will require even higher subsidies, which may be problematic. This potential aggravation thus underscores the need for policy-makers to implement effective measures (such as the examples given above) to counteract increasing demand coincidence.} 

\subsection{Methodological Contributions}
\label{sec:disc_method_contribution}


Apart from the previous findings, this paper also advances methodology related to designing electricity distribution networks. Our methodological contribution comprises the formulation and the solution of a new, generalized decision problem for planning electricity distribution networks: the \emph{\pnl}~(\pns). In more detail, the \pns extends previous decision problems \citep[\eg][]{Avella.2005} by taking into account variable degrees of demand coincidence. We provide an exact solution method for small problem instances relying on linearizations. Because of the NP-hardness, we solve the problem for larger problem instances via heuristic solution approaches. 


In the following, we discuss the benefits of our methodological advances for distribution network management. Our decision problem supports in determining long-term investment budgets. Additionally, our decision problem provides value for operational network planning tasks. When new devices with coinciding demand, such as photovoltaic systems or charging stations for electricity vehicles, are connected to a distribution network, network planners can use our \pns to stress-test existing network infrastructures. Thereby, the planners can determine bottlenecks in their networks and compute cost-effective expansions taking into account an increasing demand coincidence. Prior approaches typically assume full coincidence of consumption, either by ignoring varying coincidence factors or taking them into account only at higher layers in the networks \citep[\eg][]{Domingo.2011,Parshall.2009}, thus highlighting the importance of considering demand coincidence during planning. To this end, our results clearly show that substantial investments can be saved by considering coincidence factors at the level of network lines.

Our decision problem also suggests that the current practice in designing electricity distribution networks is suboptimal. In industry, distribution networks are commonly planned using manual procedures \citep{Gust.2017}. 
\NEW{Planning is conducted in an ad-hoc manner, \eg when new consumers or large devices (such as photovoltaic systems) need to be connected to the network. In these cases, utilities use values prescribed by industry standards for peak demands $D_i$ and coincidence factors $\gamma$, which contain a certain amount of overprovisioning \citep{kaufmann1995planung, schulz2010elektrische}.}
Some of our baselines for solving the \pns (\eg the increased network branching heuristic) mimic these manual procedures. We thereby find that such practice-oriented procedures lead to solutions that are more expensive and take significantly longer to compute in comparison to our proposed solution approach. Here, our results suggest large potential benefits when adopting our decision problem and the solution approach in practice. 
\NEW{However, in some regions, utilities are still remunerated via a cost-plus regulation scheme, which does not provide incentives for cost-efficient network investments \citep{costplus}.}

\subsection{Limitations and Alternative Solution Approaches}\label{sec:limitations}

\NEW{While the \pns models fundamental characteristics for long-term investment planning of electricity distribution networks---such as voltage drops, current carrying capacities, and radiality---the model also has limitations. Most notably, further physical and engineering-related characteristics, such as circuit protection, line losses, reactive power, and volt-var control equipment (\eg capacitor banks, on-load tap changers, and step-voltage regulators) are not included.} 

\NEW{Furthermore, the model does not cover all potential development paths of future power systems. In future scenarios where many sectors, such as transportation, heating, and industrial processes, are electrified (also referred to as ``deep electrification''; see, \eg \citet{abhyankar2017techno, gaur2020deep, wei2019electrification}), three basic scenarios may occur. First, deep electrification happens without considerable penetration of distributed energy sources (\emph{``demand extreme case''}): The electricity flows in distribution networks, peak demand levels, and coincidence of the consumption will increase. The \pns model and solution methods are capable of covering this scenario due to the coincidence factor. Second, deep electrification occurs with a very strong penetration of distributed energy sources. Each vertex of the network acts as a supplier for the power system due to excess generation (\emph{``supply extreme case''}): The \pns and its solution algorithms can be adjusted to cover this scenario by changing the direction of all flow variables, so that electricity flows from the leaf vertices towards the transformer vertex at the root of the network. Third, deep electrification occurs with a moderate penetration of distributed energy sources, where some vertices will demand energy and others will act as suppliers to the power system (\emph{``middle case''}): In this scenario, disjoint subtrees (``energy islands'') can arise, each acting as an independent network, and it must then be decided how to match demand and supply vertices. This scenario is not covered by the \pns, and different models are thus needed.} 

\NEW{Finally, coincidence factors are one concept to capture the uncertainty in peak demand (or generation) patterns for the purpose of distribution network design; there are, however, also alternative approaches. Among these approaches (see \citet[][]{EHSAN20191509, ZUBO2017} for an overview), most prominent methodologies to handle the mentioned uncertainties are stochastic optimization \citep[\eg][]{Ding2021, Fan2020, Jooshaki2020, munozdelgado2016}, robust optimization \citep[\eg][]{ZDRAVESKI2023109043, SOUZA2011527}, and risk-based approaches \citep[\eg][]{samper2013}. These approaches generate a joint probability distribution of demand (or generators) that is then used as input for optimizing the network design. The objective functions either optimize expected values or quantiles for making the decisions more robust to extreme scenarios. Both kinds of objectives typically require considerable input data to generate the distributions.

Accordingly, network design based on coincidence factors has two attractive properties: first and foremost, no extensive measurement data on the loads are required. Unlike in transmission networks \citep[\eg][]{MarkleHu.2019}, low-voltage distribution networks lack measuring systems, so that the network state, and in particular, the joint distribution of the demand is generally not available for larger sets of networks---which is a major drawback of the stochastic optimization techniques \citep{Dickert.2010}. Yet, coincidence factors incorporate the uncertainties inherent in demand (or generation) in a deterministic model. Thus, they enable the use of deterministic optimization approaches, which are computationally much more tractable than their stochastic counterparts.}

\let\oldbibliography\thebibliography
\renewcommand{\thebibliography}[1]{%
\oldbibliography{#1}%
\setlength{\itemsep}{0pt}
}

\SingleSpacedXI
\bibliographystyle{informs2014} 
\bibliography{literature, grid_planning_gg,literature_revision} 
\OneAndAHalfSpacedXI







\ECSwitch


\ECHead{Online Supplement}
\renewcommand{\thesection}{\Alph{section}}
\renewcommand{\theequation}{\thesection.\arabic{equation}}

\section{Proofs of Statements}
\label{sec:proofs}

\subsection{NP-hardness}\label{sec:NP_hardness}
We derive NP-hardness by reduction. We show that the \pns is a generalized form of the problem in \cite{Brimberg.2003}, which is known to be NP-hard. More precisely, we show that the problem in \cite{Brimberg.2003} is a special case of the \pns with $U_{\mathrm{crit}} = 0$ and a uniform coincidence factor $\gamma \equiv 1$.

By setting $U_{\mathrm{crit}} = 0$, we can ignore the constraint for the voltage drops in \Crefrange{eq:constr_voltage_1}{eq:constr_voltage_3}. We further set $\gamma \equiv 1$, which yields 
\begin{align}
	& \text{min} && \sum\limits_{(i, j)^k \in E} x_{ij}^k \, [l_{ij} c_{\mathrm{c}} + l_{ij} c_{\mathrm{m}}\, a_{ij}^k] 	
	\\
  & \text{s. t.} && \sum \limits_{k \in \{1, \ldots, |A|\}} x_{ij}^k a_{ij}^{k} \geq  F_{ij} \, ,	\quad \forall i, j \in \{0, \ldots, N - 1 \} \, , 
	\\
	\newsubeqblock
	\mysubeq & &&  \sum\limits_{j} F_{ji} - \sum\limits_{j} F_{ij} = D_{i}  
	\\
	\mysubeq & && \sum\limits_{i} \sum \limits_k  x_{ij}^k = 1 \, , 	\quad \forall j \in \{1, \ldots, N - 1 \}	\, . 
\end{align}
\noindent
This problem is equivalent to the problem in \cite{Brimberg.2003} with edge cost set to $l_{ij} c_{\mathrm{c}} + l_{ij} c_{\mathrm{m}}\, a_{ij}^k$. As our reduction is clearly of polynomial time, this shows that our \pns is NP-hard. 


\subsection{Proof of \Cref{pro:ratio}}

\begin{repeattheorem}[\Cref{pro:ratio}.]
For any given network layout and for continuous capacities (\ie $A = \mathbb{R}^+$), the capacities that minimize the cost fulfill
	\begin{align}
		\frac{a_{ij}^2}{a_{mn}^2} = \frac{F_{ij}}{F_{mn}} \quad \text{or} \quad \frac{a_{ij}}{a_{mn}} = \sqrt{\frac{F_{ij}}{F_{mn}}} 
	\end{align}
\REVTWO{for all combinations of edges}  $(i,j)$ and $(m,n)$ in the same path $p \in P$ and if \Cref{eq:constr_voltage_2} is binding.
\end{repeattheorem}
\noindent
\proof{Proof.}
We assume that the set of possible capacities is continuous and all capacities can take up any real value. As a first step, we consider a very simple network consisting of three vertices (with $i = 0, 1, 2$) and two edges. Later, we expand this to networks of arbitrary length and branching. In the first step, one edge connects vertex 1 to the source. Vertex 2 is connected to vertex 1 by a second edge. Using the objective function in \Cref{eq:obj_fun}, we can set up the cost function for this network as
\begin{align}
C = l_{01} \, c_\mathrm{c} + l_{01} \, c_\mathrm{m} \, a_{01} + l_{12} \, c_\mathrm{c} + l_{12} \, c_\mathrm{m} \, a_{12}
\, .
	\label{eq:branch_obj_fun}
\end{align}
For a given solution to the \pns, we derive the accumulated voltage drop from \Cref{eq:new_voltage}, which yields
\begin{align}
Q =  l_{01} \frac{F_{01}}{a_{01}} + l_{12} \frac{F_{12}}{a_{12}}
\, .
	\label{eq:branch_constr_voltage}
\end{align}
To minimize the cost function in \Cref{eq:branch_obj_fun}, we derive the Lagrangian for this problem, which yields
\begin{align}
\mathcal{L} = c_\mathrm{c}  (l_{01} + l_{12}) + c_\mathrm{m}  (l_{01} \, a_{01} + l_{12} \, a_{12}) + \lambda   \left(l_{01} \frac{F_{01}}{a_{01}} + l_{12} \frac{F_{12}}{a_{12}} - Q  \right) 
 ,
\label{eq:branch_lagrange}
\end{align}
\noindent
where $\lambda$ is the Lagrange multiplier to include the voltage drop from \Cref{eq:branch_constr_voltage}. We then take the partial derivatives with respect to $a_{01}$, $a_{12}$ and $\lambda$. By setting them to zero, we arrive at the following system of equations:
\begin{alignat}{2}
	 \frac{\partial \mathcal{L}}{\partial a_{01}} 	& = l_{01} \, c_\mathrm{m} - \lambda \, l_{01} \frac{F_{01}}{(a_{01})^2} 	&&= 0 \, , \label{eq:branch_2} 
	\\
	 \frac{\partial \mathcal{L}}{\partial a_{12}} 	& = l_{12} \, c_\mathrm{m} - \lambda \, l_{12} \frac{F_{12}}{(a_{12})^2} 	&&= 0 \, , \quad \text{and} \label{eq:branch_3}
	\\
	 \frac{\partial \mathcal{L}}{\partial \lambda} 	& =  \frac{l_{01} F_{01}}{a_{01}} + \frac{l_{12} \, F_{12}}{a_{12}} - Q 	&&= 0 \, \label{eq:branch_4} .
\end{alignat}
\noindent
In \Cref{eq:branch_2,eq:branch_3}, the lengths cancel out, and the two formulas can be rewritten to
\begin{align}
	 c_\mathrm{m} - \lambda  \frac{F_{01}}{a_{01}^2} & = 0 \, , \quad \text{and} \label{eq:branch_2_new} 
	\\
	 c_\mathrm{m} - \lambda  \frac{F_{12}}{a_{12}^2} & = 0 \, . \label{eq:branch_3_new}
\end{align}
\noindent
From \Cref{eq:branch_2_new,eq:branch_3_new}, a generalized formula for paths of arbitrary length can be derived. The generalized formula is
\begin{align}
c_\mathrm{m} - \lambda \frac{F_{ij}}{a_{ij}^2} = 0
\, ,
	\label{eq:branch_generalized}
\end{align}
which can be rewritten to
\begin{align}
a_{ij}^2 = \frac{\lambda}{c_\mathrm{m}}  F_{ij}
\, .
	\label{eq:branch_generalized_2nd}
\end{align}
\noindent
The square of the capacity of an edge $a_{ij}^2$ is proportional to the flow $F_{ij}$. For any two edges $(i,j)^k$ and $(m,n)^k$ of the same path $p \in P$, we find
\begin{align}
\frac{a_{ij}^2}{a_{mn}^2} = \frac{F_{ij}}{F_{mn}} \, .
\end{align}
\noindent
This concludes our proof. $ \square $
\endproof

\subsection{Proof of \Cref{pro:min}}

\begin{repeattheorem}[\Cref{pro:min}.]
\textit{For $D_i \to 0$ (\REVTWO{or, alternatively, $D_i > 0$ and  $a_{\mathrm{min}} = \underset{k}{\min} \; a_{ij}^k$ satisfying the flow and voltage drop constraints}}), the MST layout is the optimal solution to the \pns.
\end{repeattheorem}
\proof{Proof.}
When $D_i \to 0$, we find that $F_{ij} \to 0$. Consequently, the constraints for both line sizing (\Cref{eq:current}) and voltage drops (\Cref{eq:constr_voltage_2}) are fulfilled for any choice of $x_{ij}^k$. (Analogously, the same holds true if the smallest line capacity, $a_{\mathrm{min}}$, is sufficiently large in comparison to the flows.) In these cases, all capacities can be set to the lowest possible value $a_{\mathrm{min}} = \underset{k}{\min} \; a_{ij}^k$. Then, the objective function in \Cref{eq:obj_fun} simplifies to $\min \; \sum\limits_{i, j} x_{ij} l_{ij} [c_{\mathrm{c}} + c_{\mathrm{m}}a_{\mathrm{min}}]$. The term $[c_{\mathrm{c}} + c_{\mathrm{m}}a_{\mathrm{min}}]$ is constant and, therefore, the objective function is identical to the objective function of the MST problem.
$ \square $
\endproof

\subsection{Proof of \Cref{pro:star}}

\begin{repeattheorem}[\Cref{pro:star}.]
If feasible solutions to the \pns exist, the star network is one of these solutions.
\end{repeattheorem}
\proof{Proof.}
We prove this proposition by contradiction. We consider a star network $\Gamma^1$ and assume that this solution violates one of the constraints in \Cref{eq:current} or \Cref{eq:constr_voltage_2} for the subtree consisting of only the edge connecting the vertex $v$ to the source $0$. We further assume that there exists an alternative solution $\Gamma^2$ not violating the constraints, and that in this solution vertex $v$ is connected to a vertex $w$ other than $0$. Without loss of generality, we assume that all capacities in both the star layout and the alternative layout are set to the maximum value $a_{\mathrm{max}} = \underset{k}{\max} \, a_{ij}^k$.

We distinguish two cases. First, we consider the line sizing constraint in \Cref{eq:current}. It is obvious that the line sizing constraint cannot be the reason why $\Gamma^2$ is feasible and $\Gamma^1$ is not, since $F_{uw}^{2} > F_{0v}^{1} = D_v$ and $D_i \in \mathbb{R}^+$, \ie the flows on the edges in the star network are minimal. It should be noted, that we require the coincidence factor to have a form such that adding demands to an edge always increases the flows, \ie $\tfrac{\gamma(|\Gamma_j| + 1)}{\gamma(|\Gamma_j|)} > \tfrac{|\Gamma_j|}{|\Gamma_j| + 1}$. This is the case for all forms present in the literature \citepappendix{Dickert.2010}. Second, we focus on the constraint for the voltage drops in \Cref{eq:constr_voltage_2}. By use of the triangle inequality, we show that the total length of all edges from the source to the vertex in $\Gamma^2$ is $L^2 \geq l_{0v} = L^1$. Therefore, \Cref{eq:constr_voltage_2} must also have been fulfilled in the star network $\Gamma^1$. This contradicts our assumption and concludes our proof.
$ \square $
\endproof

\MOVED{
\subsection{Error bounds of the PECA heuristic} \label{sec:proofs_PECA}

In the following, we provide theoretical guarantees for the accuracy of the approximate solution to the capacity optimization problem of the PECA heuristic. To this end, we provide upper bounds for the error made by the PECA heuristic. The formulas provided in \Cref{pro:PECA,pro:PECa^2} refer to the maximum error made when optimizing an edge pair in \cref{line:peca_opt} of the heuristic. The error for the entire network can be calculated by summing over all edge pairs. For this reason, we introduce the following notation. Let $\Delta_c$ be the cost difference between the optimal cost $c^*$ for an edge pair and the cost for the same edge pair as determined by the PECA heuristic $c^{\mathrm{PECA}}$, \ie $\Delta_c = c^{\mathrm{PECA}} - c^*$. Furthermore, let $\Delta_a$ be the maximum difference between any two consecutive capacities $a_{ij}^k, a_{ij}^{k+1} \in A$, and $\Delta_l$ the difference in the length of the two edges. 

\vspace{0.15cm}
\begin{proposition}
\label{pro:PECA}
For the network determined by the PECA heuristic, the error in cost for each edge pair is bounded by $\Delta_c \leq c_m \, \Delta_a \, \Delta_l$.
\end{proposition}
}

\proof{Proof.}
Without loss of generality, we assume that the capacities $a_{ij}^k \in A$ are ordered, with the first element $a_{ij}^1$ being the smallest. For now, we assume that the capacities in $A$ are equally spaced, \ie that $a_{ij}^{k+1} - a_{ij}^k = \Delta_a,$ for all $k$. In the case of non-equally spaced values, we define $\Delta_a = \underset{k}{\max} \{a_{ij}^{k+1} - a_{ij}^k\}$ as the maximum difference between any two subsequent values in $A$. Let us consider a pair of edges $x$ and $y$ for which the capacity is subject to optimization using the PECA heuristic. Let $l_x$ and $l_y$ be the lengths of these edges. 

We assume that the the optimal solution to the capacity optimization problem differs from the one found by the PECA heuristic. Let $a^x$ and $a^y$ be the capacities of the optimal solution. Without loss of generality, we assume that $a^x > a^y$, and hence, $x > y$. Both $a^x$ and $a^y$ cannot be identical, since, otherwise, the PECA heuristic would have identified them as the optimal solution. Using the cost function in \Cref{eq:new_objective}, we can write down the optimal combined cost $c^*$ for the two edges, which gives
\begin{align}
c^* = c_m \, l_x \, a^x + c_m \, l_y \, a^y \, .
\label{eq:PECA_opt}
\end{align}

Owing to the discrete nature of our problem, the PECA heuristic might determine capacities different from the optimal values $a^x$ and $a^y$ due to rounding. One of the edges must then have a capacity larger than in the optimal case, whereas the other must be smaller. Without loss of generality, we can assume that the heuristic has chosen the pair $a^{x+1}$ and $a^{y-1}$ as we obtain the same result if the pair $a^{x-1}$ and $a^{y+1}$ is chosen. Furthermore, we arrive at the same result if the capacities determined by the PECA heuristic differ from the optimal solution by more than one index.

The cost determined by the PECA heuristic then is
\begin{align}
c = c_m l_x a^{x+1} + c_m l_y a^{y-1} \, .
\label{eq:PECA_cost}
\end{align}
The difference between this cost and the optimal cost is obtained by subtracting \Cref{eq:PECA_opt} from \Cref{eq:PECA_cost}. This gives 
\begin{align}
	 \Delta_c = c - c^* & = c_m [l_x \, a^{x+1} + l_y \, a^{y-1} - l_x \, a^x - l_y \, a^y]
	\\
												& = c_m [(a^{x+1} - a^x) l_x - (a^y - a^{y-1}) l_y]
	\\
												& = c_m \Delta_a \Delta_l \, \label{eq:delta_c}.
\end{align}

$ \square $
\endproof

\noindent
\MOVED{As a side observation, \Cref{pro:PECA} implies that, for networks with roughly equal edge lengths, the error approaches zero (\Cref{pro:PECA_zero}).
\vspace{0.15cm}

\begin{corollary}
\label{pro:PECA_zero}
\textit{If the length difference of an edge pair approaches zero, the cost error approaches zero, \ie $\Delta_c \xrightarrow{\Delta_l \to 0} 0$. Therefore, if all edges in the network are of equal length, the error for the entire network approaches zero.}
\end{corollary}
\noindent
We now provide an upper bound for the error that is independent of the edge lengths. Let $a_{ij}$ be the selected capacity of the edge closer to the source.

\begin{proposition}
\label{pro:PECa^2}
For any edge pair optimized, the relative cost error is $\frac{\Delta_c}{c^*} \leq 1 / \left(\frac{a_{ij}}{\Delta_a} - 1 \right)$.
\end{proposition}
}
\proof{Proof.}
We use the same notation as in the proof of \Cref{pro:PECA}. We denote the length and capacity of the edge closer to the source by the index $x$ and of the other edge in the edge pair by $y$. Let $a^x$ and $a^y$ denote the capacities of the optimal solution to the capacity optimization problem. Without loss of generality, we assume $l_x = l_y + \Delta_l$. We obtain
\begin{align}
	 \frac{\Delta_c}{c^*} \leq \frac{c_m \Delta_a \Delta_l}{c_m \, l_x \, a^{x} + c_m \, l_y \, a^{y}} = \frac{\Delta_a \, \Delta_l}{(l_y + \Delta_l) a^{x} + l_y \, a^{y}} \, . \label{eq:PECA_bound_1} 
\end{align}
\noindent
We substitute the optimal capacity $a^{x}$ in the denominator by the one determined by the PECA heuristic using $a^{x} = a^{x-1}$ and obtain
\begin{align}
	 \frac{\Delta_c}{c^*} \leq \frac{\Delta_a \, \Delta_l}{(l_y + \Delta_l) a^{x-1} + l_y \, a^{y}}  \, . \label{eq:PECA_bound} 
\end{align}
We now analyze the border cases of this expression. In the best case, we obtain 
\begin{align}
	 \lim_{\Delta_l \to 0} \frac{\Delta_c}{c^*} = 0 \, .
\end{align}
In the worst case, the upper bound for the relative cost difference is
\begin{align}
	 \lim_{\Delta_l \to \infty} \frac{\Delta_c}{c^*} \leq \frac{\Delta_a}{a^{x-1}} = \frac{\Delta_a}{a^{x} - \Delta_a} = \frac{1}{\frac{a_{ij}}{\Delta_a} - 1} \, . \label{eq:PECA_worst_case}
\end{align}
$ \square $
\endproof

\noindent
From \Cref{pro:PECa^2}, it follows that the error decreases the smaller the difference between $a_{ij}^{k-1}$ and $a_{ij}^k$ is.

\section{Linearization of the Problem Formulation} \label{sec:lin}

Below, we present a linearization of the \pns. The linearization of the flow constraint in \Cref{eq:flow} is the most complex, as it requires first a piecewise linearization of the coincidence factor $\gamma(|\Gamma_j|)$, which then yields a cubic nonlinearity. We discuss this linearization in \Cref{sec:lin_complete}. Linearization of the quadratic nonlinearity in the voltage drop constraint in \Cref{eq:constr_voltage_1} is required to derive upper and lower bounds for the problem. We discuss this linearization in \Cref{sec:alt_lin}. The remaining linearizations are discussed in \Cref{sec:qua}.

\subsection{Linearization of Flow Constraint} \label{sec:lin_complete}

The first step in linearizing \Cref{eq:flow} is a piecewise linearization of the coincidence factor $\gamma(|\Gamma_j|)$. This is achieved by introducing a binary variable $w_{jn}$, together with the constraints
\begin{align}
w_{jn} &=
    \begin{cases}
      1, & \text{if}\ |\Gamma_j| = n \; , \\
      0, & \text{otherwise} \; ,
    \end{cases}
\quad \forall j, n \in V \; ,
\\
\sum\limits_n w_{jn} &= 1 \; , \quad \forall j \in V \; .
\end{align}
Next, we introduce a variable $\gamma_j$, with 
\begin{align}
\gamma_j = \sum\limits_n \gamma(n) w_{jn} \; , \quad \forall j \in V \; .
\end{align}
After this linearization of $\gamma(|\Gamma_j|)$, there is a cubic nonlinearity in \Cref{eq:flow} because of the product of three variables, namely $x_{ij}^k$, $w_{jn}$, and $\overline{D}_j$.

We introduce a new variable $z_{ijn}^k$, representing the product $z_{ijn}^k \define x_{ij}^k w_{jn}$. This is done by using the inequalities
\begin{align}
\newsubeqblock
\mysubeq z_{ijn}^k &\leq x_{ij}^k \; , 							&&\forall i, j, n \in V \, , \; \forall k \in \{1, \ldots, |A|\}  \; ,
\\
\mysubeq z_{ijn}^k &\leq w_{jn} \; ,									&&\forall i, j, n \in V \, , \; \forall k \in \{1, \ldots, |A|\}  \; ,
\\
\mysubeq z_{ijn}^k &\geq x_{ij}^k + w_{jn} -1 \; , 	&&\forall i, j, n \in V \, , \; \forall k \in \{1, \ldots, |A|\}  \; .
\end{align}
Analogously, we introduce a second variable $z_{ijn}^{'k}$ for the product $z_{ijn}^{'k} \define x_{ij}^k w_{in}$. 

In the last step, we are left with the quadratic nonlinearity resulting from the products $z_{ijn}^{k} \overline{D}_j$ and $z_{ijn}^{'k} \overline{D}_j$. This can be resolved via the Big~M method. We introduce two new variables $\Delta_{ijn}^k \define z_{ijn}^k \overline{D}_j$ and $\Delta_{ijn}^{'k} \define z_{ijn}^{'k} \overline{D}_j$, together with
\begin{align}
\newsubeqblock
\mysubeq \Delta_{ijn}^k &\leq z_{ijn}^k M \; , &&\forall i, j, n \in V \, , \; \forall k \in \{1, \ldots, |A|\} \; , 
\\
\mysubeq \Delta_{ijn}^k &\geq 0 \; ,					&&\forall i, j, n \in V \, , \; \forall k \in \{1, \ldots, |A|\} \; , 
\\
\mysubeq \Delta_{ijn}^k &\leq \overline{D}_j \; , &&\forall i, j, n \in V \, , \; \forall k \in \{1, \ldots, |A|\} \; , 
\\
\mysubeq \Delta_{ijn}^k &\geq \overline{D}_j - (1- z_{ijn}^k) M \; , &&\forall i, j, n \in V \, , \; \forall k \in \{1, \ldots, |A|\} \; , 
\end{align}
and analogous constraints for $\Delta_{ijn}^{'k}$.

As a result, we arrive at the linearized version of \Cref{eq:flow}, which reads
\begin{align}
\sum\limits_{j} F_{ji} - \sum\limits_{j} F_{ij} = \gamma_i D_i - \sum\limits_{j} \sum \limits_k \left( \sum\limits_n \gamma(n) \Delta_{ijn}^k - \sum\limits_n \gamma(n) \Delta_{ijn}^{'k} \right)  \, , \quad		 \forall i \in V \setminus \{0\}	\, .
\end{align}

\subsection{Linearization of Voltage Drop Constraint} \label{sec:alt_lin}

In the following, we linearize the quadratic constraint representing Ohm's law in \Cref{eq:constr_voltage_1} using the Big M notation. In doing so, we follow the approach of \cite{Avella.2005} and replace \Cref{eq:constr_voltage_1} with two linear inequalities. This yields
\begin{align}
\frac{a_{ij}^k}{l_{ij}} (U_i - U_j) &\leq F_{ij} + (1 - x_{ij}^k) M , \, & \forall	i, j \in V \, , \, \forall k \in \{1, \ldots, |A|\} , \label{eq:M_1}	
\\
\frac{a_{ij}^k}{l_{ij}} (U_i - U_j) &\geq F_{ij} - (1 - x_{ij}^k) M , \, & \forall	i, j \in V \, , \, \forall k \in \{1, \ldots, |A|\} , \label{eq:M_2}
\end{align}
where $M$ is a suitable large number. If an edge is selected (\ie if $x_{ij}^k = 1$), the combination of \Crefrange{eq:M_1}{eq:M_2} is equivalent to \Cref{eq:constr_voltage_1}. If  $x_{ij}^k = 0$, the equations become redundant.

\subsection{Linearization of Remaining Nonlinearities} \label{sec:qua}

We linearize \Cref{eq:gamma_i,eq:d_bar_i} by introducing two new variables $g_{ij}^k$ and $d_{ij}^k$, representing the products $g_{ij}^k \define x_{ij}^k \, |\Gamma_j|$ and $d_{ij}^k \define x_{ij}^k \, \overline{D}_j$. These products can be written down in a linear way by using the Big M method. This yields
\begin{align}
\newsubeqblock
\mysubeq g_{ij}^k &\leq x_{ij}^k M \; , &&\forall i, j \in V \setminus \{0\} \, , \; \forall k \in \{1, \ldots, |A|\} \; ,
\\
\mysubeq g_{ij}^k &\geq 0 \; , &&\forall i, j \in V \setminus \{0\} \, , \; \forall k \in \{1, \ldots, |A|\} \; , 
\\
\mysubeq g_{ij}^k &\leq |\Gamma_j| \; , &&\forall i, j \in V \setminus \{0\} \, , \; \forall k \in \{1, \ldots, |A|\} \; ,
\\
\mysubeq g_{ij}^k &\geq |\Gamma_j| - (1- x_{ij}^k) M \; , &&\forall i, j \in V \setminus \{0\} \, , \; \forall k \in \{1, \ldots, |A|\} \; ,
\end{align}
as well as
\begin{align}
\newsubeqblock
\mysubeq d_{ij}^k &\leq x_{ij}^k M \; ,  		&&\forall i, j \in V \, , \; \forall k \in \{1, \ldots, |A|\} \; ,
\\
\mysubeq d_{ij}^k &\geq 0 \; , 				 			&&\forall i, j \in V \, ,  \; \forall k \in \{1, \ldots, |A|\} \; , 
\\
\mysubeq d_{ij}^k &\leq \overline{D}_j \; , &&\forall i, j \in V \, , \; \forall k \in \{1, \ldots, |A|\} \; ,
\\
\mysubeq d_{ij}^k &\geq \overline{D}_j - (1- x_{ij}^k) M \; ,  &&\forall i, j \in V \, , \; \forall k \in \{1, \ldots, |A|\} \; .
\end{align}
$M$ is a suitable large number. With these equations, the linearized forms of \Cref{eq:gamma_i,eq:d_bar_i} read
\begin{align}
\newsubeqblock
	\mysubeq  \abs{\Gamma_i} &= 1 + \sum \limits_{j} \sum \limits_{k} g_{ij}^k  \, , &&\forall i \in V	\, ,  \label{eq:gamma_i_lin}
	\\
	\mysubeq  \overline{D}_i &= D_i + \sum \limits_{j} \sum \limits_{k} d_{ij}^k \, , &&\forall i \in V 	\, .  \label{eq:d_bar_i_lin}
\end{align}


\section{Performance Evaluation Against Alternative Solution Approaches} \label{sec:evaluation_methods}

\subsection{Complete \pns}

In this section, we present alternative solution approaches which are used to evaluate the performance of the Tabu Search metaheuristic. \NEW{These approaches include an alternative model formulation of the \pns based on undiscounted flows and its solution using the Gurobi MIP solver}, as well as two heuristics, namely variable neighborhood search~(VNS) heuristic and the increased network branching (INB) heuristic. The latter heuristic is based on actual network reinforcement practice. The approaches are presented in \crefrange{sec:commodities_model}{sec:INB} and their performance is evaluated in \cref{sec:exp_setup}. 

\subsubsection{\NEW{\pns with Undiscounted Flows.}} \label{sec:commodities_model}


\NEW{
We present an alternative model formulation of the \pns based on undiscounted flows. Instead of discounting the flows resulting from electricity demands $D$ with the coincidence factor (this is achieved by \cref{eq:flow} in the \pns), and using the resulting $F_{ij}$ as flow variable, this alternative model uses undiscounted demand $q_{ij}$ as primary flow variable. This modification requires the introduction of a second flow variable, $y_{ij}$, that keeps track of the number of vertices that are located downstream of each edge $(i,j)^k$. \Cref{tbl:notation_commodities} summarizes the additional nomenclature.

\begin{table}[ht]
\centering
\caption{\NEW{Additional nomenclature for the \pns with Undiscounted Flows.} \label{tbl:notation_commodities}}
{\resizebox{.7\columnwidth}{!}{%
\begin{tabular}{lll}
\toprule
\textbf{Symbol} & \textbf{Description} & \textbf{Unit/range} \\
\midrule
$ q_{ij}$ & Demand served via edge $(i,j)^k$ & $q_{ij} \in \mathbb{R}_0^+ $\\
$ y_{ij}$ & Number of vertices connected via edge $(i,j)^k$ & $0 \leq y_{ij} \leq N-1 $\\
$ M $ & Big M  & $M \in \mathbb{R}^+ $ \\
\bottomrule
\end{tabular}}
}
{}
\end{table}

As in the \pns, the source location is defined as vertex $0$, and the decision variable $x_{ij}^k \in \left\{ 0, 1 \right\}$ indicates whether an edge of type $k$ from vertex $i$ to vertex $j$ should be built. The cumulative demand and cumulative number of vertices at edge $(i,j)^k$ --- \ie the sum of demands of all vertices and the number of vertices connected up to vertex $j$ (including vertex $j$) --- are denoted by $q_{ij} $ and $y_{ij}$, respectively.

The \pns with undiscounted flows is then given by
\begin{align}
	& \text{min} && \sum\limits_{(i, j)^k \in E} x_{ij}^k \, [l_{ij} c_{\mathrm{c}} + l_{ij} c_{\mathrm{m}}\, a_{ij}^k] 	\label{eq:obj_fun_commodities}	
	\\
  & \text{s. t.} && \sum \limits_{k \in \{1, \ldots, |C|\}} x_{ij}^k c_{ij}^{k} \geq  \gamma \left( y_{ij} \right)q_{ij} \, ,	\quad &&\forall (i, j) \in E \, ,   \label{eq:current_commodities}
	\\
        \newsubeqblock
	\mysubeq & &&  \sum\limits_{j} q_{ji} - \sum\limits_{j} q_{ij} = D_i \, , \quad &&\forall i \in V \setminus \{ 0 \}	 \, ,  \label{eq:demand_flow_commodities} 
	\\
        \mysubeq & &&  \sum\limits_{j} q_{0j} = \sum\limits_{i} D_i \, , \label{eq:demand_flow_root_commodities}
        \\
        \mysubeq & &&  q_{ij} \leq \sum\limits_{k} x_{ij}^k M \, , \quad  &&\forall (i, j) \in E \, , \label{eq:demand_flow_capacity_commodities}        
        \\
	\mysubeq & &&  \sum\limits_{j} y_{ji} - \sum\limits_{j} y_{ij} = 1 \, , \quad &&\forall i \in V \setminus \{ 0 \}	 \, ,  \label{eq:nnodes_flow_commodities} 
	\\
        \mysubeq & &&  \sum\limits_{j} y_{0j} = N-1 \, , \label{eq:nnodes_flow_root_commodities}
        \\
        \mysubeq & &&  y_{ij} \leq \sum\limits_{k} x_{ij}^k M \, , \quad  &&\forall (i, j) \in E \, , \label{eq:nnodes_flow_capacity_commodities}
	\\
	\mysubeq & && \sum\limits_{i} \sum \limits_k  x_{ij}^k = 1 \, , 	\quad &&\forall j \in V \setminus \{ 0 \}	\, ,  \label{eq:tree_commodities}
	\\
	\newsubeqblock
	\mysubeq & && \sum\limits_{k} x_{ij}^k \frac{a_{ij}^k}{l_{ij}} (U_i - U_j) = \gamma \left( y_{ij} \right)q_{ij} \, , \quad  &&\forall (i, j) \in E \, ,     \label{eq:constr_voltage_1_commodities}
	\\
	\mysubeq & && U_i \geq U_{\mathrm{crit}} \, , \quad &&\forall  i \in V \, ,     \label{eq:constr_voltage_2_commodities}
		\\
	\mysubeq & && U_0 = U  \, .     \label{eq:constr_voltage_3_commodities} 
 \\
   &  && x_{ij}^k \in \{0,1\} \, ,	\quad && \NEW{\forall (i, j)^k \in E \, .}   \label{eq:binary_commodities}
\end{align}

\noindent
The objective in \Cref{eq:obj_fun_commodities} is the same as that of the \pns. The line sizing constraint in \Cref{eq:current_commodities} requires the edge $(i, j)^k$ to be sufficiently large to support the cumulative flow discounted with the coincidence factor $\gamma \left( y_{ij} \right)q_{ij}$. \Crefrange{eq:demand_flow_commodities}{eq:tree_commodities} define the flows and ensure radial layout and connectivity. \Cref{eq:demand_flow_commodities} and \cref{eq:demand_flow_root_commodities} ensure flow conservation for the undiscounted flows.  \Cref{eq:nnodes_flow_commodities} and \cref{eq:nnodes_flow_root_commodities} serve the same purpose for the second flow variable $y_ij$. \Cref{eq:demand_flow_capacity_commodities,eq:nnodes_flow_capacity_commodities} force both flow variables to zero if no edge from vertex $i$ to vertex $j$ is built. Since $ y_{ij}$ accumulates the number of vertices served by node $j$ and nodes further downstream, it enables the magnitude of the coincidence factor $\gamma(y_{ij})$ to be determined, as necessary for \Cref{eq:current_commodities,eq:constr_voltage_1_commodities}. Conceptually, \cref{eq:constr_voltage_1_commodities} represents the same voltage constraint as that of the \pns, but the undiscounted flows $q_{ij}$ still need to be adjusted with the coincidence factor on the right-hand side. The remaining constraints are kept unchanged.}


\NEW{
Note that \Cref{eq:current_commodities,eq:constr_voltage_1_commodities} need to be linearized in this formulation. For \Cref{eq:current_commodities} we used a similar approach to the one presented in \Cref{sec:lin_complete} but we replace the binary variable $w_{jn}$ with $w_{ijn}$ such as

\begin{align}
w_{ijn} &=
    \begin{cases}
      1, & \text{if}\ y_{ij} = n \; , \\
      0, & \text{otherwise} \; .
    \end{cases}
\quad \forall i, j, n \in V \; ,
\end{align}

Then, we replace $\gamma \left( y_{ij} \right)$ with $\gamma_{ij}$ as
\begin{align}
\gamma_{ij} = \sum\limits_n \gamma(n) w_{ijn} \; , \quad \forall i, j \in V \; .
\end{align}

Next, we use $\Delta_{ijn}$, representing the product $\Delta_{ijn} \define w_{ijn} q_{ij}$ with the following constraints

\begin{align}
\newsubeqblock
\mysubeq \Delta_{ijn} &\leq w_{ijn} M \; , &&\forall i, j, n \in V \, , 
\\
\mysubeq \Delta_{ijn} &\geq 0 \; ,					&&\forall i, j, n \in V \, , 
\\
\mysubeq \Delta_{ijn} &\leq q_{ij} \; , &&\forall i, j, n \in V \, ,
\\
\mysubeq \Delta_{ijn} &\geq q_{ij} - (1- w_{ijn}) M \; , &&\forall i, j, n \in V \, . 
\end{align}

Together, this leads to the linearized version of \Cref{eq:current_commodities}, which reads

\begin{align}
\sum \limits_{k \in \{1, \ldots, |C|\}} x_{ij}^k c_{ij}^{k} \geq  \sum\limits_n \gamma(n) \Delta_{ijn} \, ,	\quad &&\forall (i, j) \in E \, .
\end{align}

For \Cref{eq:constr_voltage_1_commodities}, we start linearizing the quadratic left-hand side of the constraint with the following two linear inequalities (as described in \Cref{sec:alt_lin}),

\begin{align}
\frac{a_{ij}^k}{l_{ij}} (U_i - U_j) &\leq \gamma \left( y_{ij} \right)q_{ij} + (1 - x_{ij}^k) M , \, & \forall	i, j \in V \, , \, \forall k \in \{1, \ldots, |A|\} ,
\\
\frac{a_{ij}^k}{l_{ij}} (U_i - U_j) &\geq \gamma \left( y_{ij} \right)q_{ij} - (1 - x_{ij}^k) M , \, & \forall	i, j \in V \, , \, \forall k \in \{1, \ldots, |A|\} .
\end{align}

Then, we linearize the quadratic term on the right-hand side using the approach described above. This yields

\begin{align}
\frac{a_{ij}^k}{l_{ij}} (U_i - U_j) &\leq \sum\limits_n \gamma(n) \Delta_{ijn} + (1 - x_{ij}^k) M , \, & \forall	i, j \in V \, , \, \forall k \in \{1, \ldots, |A|\} , \label{eq:constr_voltage_1_linear_1}	
\\
\frac{a_{ij}^k}{l_{ij}} (U_i - U_j) &\geq \sum\limits_n \gamma(n) \Delta_{ijn} - (1 - x_{ij}^k) M , \, & \forall	i, j \in V \, , \, \forall k \in \{1, \ldots, |A|\} . \label{eq:constr_voltage_1_linear_2}
\end{align}
  
}

\subsubsection{Variable Neighborhood Search Metaheuristic.}
\label{sec:VNS}

The idea behind VNS is to change neighborhoods in order to find a better solution, as opposed to local search methods that do not use several neighborhoods within one method \citepappendix[\cf][]{Hansen.2014, Hansen.2019, Mladenovic.1997}. For the change in neighborhoods, a metric for the distance between solutions needs to be introduced. In our case, we define the distance between two layouts by their difference in edges. For example, a certain layout $A$ has distance 1 to another layout $B$ if it can be reached by adding just one edge to $B$ and deleting another. We then say that that $A$ is part of the ``1-neighborhood'' of $B$.

VNS has two main components, which are (a) shaking (\ie the change of neighborhoods) and (b) local search. The pseudocode is provided in \Cref{alg:VNS}. In \crefrange{line:VNS_1}{line:initial_VNS}, the heuristic determines capacities and initial cost and saves the initial layout. The VNS procedure starts in \cref{line:VNS_for} and runs for $s_{\mathrm{VNS}}$ iterations. In \cref{line:VNS_distance}, the distance $d$ is set to 1. In the first iteration, the algorithm thus starts exploring the 1-neighborhood of the initial solution. The algorithm explores neighborhoods in a distance of up to $d_\mathrm{max}$ (\cref{line:VNS_dmax}). \Cref{line:VNS_shaking_start,line:VNS_shaking_end} state the shaking procedure. We use intensified shaking, where, instead of conventional shaking by drawing an arbitrary neighboring layout, a more strategic procedure is applied \citepappendix[see, \eg][]{Brimberg.2003, Hansen.2014}. For this purpose, we use the randomized network reconfiguration method with just one iteration (\ie $s_\mathrm{max}=1$) to choose one edge that is to be added at random and then find the best edge to be removed. In \cref{line:VNS_local}, the local search is performed. The cost for the best layout found is determined in \crefrange{line:VNS_cost_start}{line:VNS_cost_end}. If the cost is smaller than the current best solution, the algorithm saves the new best solution in \crefrange{line:VNS_if}{line:VNS_reset} and proceeds with the next iteration. Otherwise, the algorithm increases $d$ to explore the next neighborhood in \cref{line:VNS_next}. At the end, the heuristic returns the cheapest network in \cref{line:return_VNS}.


	\begin{algorithm}
	\SingleSpacedXI
		\footnotesize 
		\caption{Variable Neighborhood Search} 																																																									\label{alg:VNS}
		\textbf{Input:} Network layout $\{x_{ij}\}$		
		\begin{algorithmic}[1]
			\State $\{ a_{ij} \} \gets \textsc{CapacityOptimizationMethod}(\{x_{ij}\})$																																						\label{line:VNS_1}
			\State $c^* \gets \textsc{Cost}\left(\{x_{ij}\}, \{a_{ij}\}\right)$	 						\Comment{Determine initial cost}																			\label{line:VNS_2}
			\State $X, X^* \gets \{x_{ij}\} $ 																																																										\label{line:initial_VNS}
			\For{$s \in \{ 1, \ldots, s_{\mathrm{VNS}} \}$}																																																				\label{line:VNS_for}
				\State $d \gets 1$																															\Comment{Set distance to 1}																					\label{line:VNS_distance}
				\While{$d \leq d_\mathrm{max}$}																																																											\label{line:VNS_dmax}
					\For{$d$ times}																																							\Comment{Intensified shaking}													\label{line:VNS_shaking_start}
						\State $X \gets \textsc{RandomizedNetworkReconfiguration}(\{x_{ij}\}, s_\mathrm{max}=1)$	\Comment{Draw neighboring layout}											\label{line:VNS_shaking_end}		
					\EndFor														
					\State Perform \textbf{Local Search} 																																																							\label{line:VNS_local}						
					\State $\{ a_{ij} \} \gets \textsc{CapacityOptimizationMethod}(\{x_{ij}\})$																																				\label{line:VNS_cost_start}				
					\State $c \gets \textsc{Cost}\left(\{x_{ij}\}, \{a_{ij}\}\right)$							\Comment{Determine new cost}																				\label{line:VNS_cost_end}
					\If{$c \leq c^*$} 																																																																\label{line:VNS_if}	
						\State $X, X^* \gets \{x_{ij}\}$ and $c^* \gets c$ 													\Comment{Update best solution if cost has improved}	
						\State \textbf{break}																												\Comment{Return to \cref{line:VNS_distance}}												\label{line:VNS_reset}	
					\Else {} $d \gets d + 1$ and $\{x_{ij}\} \gets X$															\Comment{Increase distance to explore next neighborhood}						\label{line:VNS_next}					
					\EndIf		
				\EndWhile
			\EndFor
		\State \textbf{return} $X^*$ 																																																														\label{line:return_VNS}
		\end{algorithmic}	
	\end{algorithm}

Our problem allows edges between any two vertices. This makes the objective function very sensitive to the shaking procedure, \ie a wrong choice of an edge to be added might strongly deteriorate the objective value regardless of which edge is subsequently removed. Therefore, in our experiments, we modify the VNS algorithm to achieve better results. This is done by further intensifying the neighborhood change via shaking in \cref{line:VNS_shaking_start,line:VNS_shaking_end}. We find that the algorithm yields better computational results if one further adaptation is made to the shaking procedure described in \cref{line:VNS_shaking_start,line:VNS_shaking_end}. Instead of shaking only once (as described in \Cref{alg:VNS}), we repeat \cref{line:VNS_shaking_end} five times and then select the best solution for the neighborhood change.

\subsubsection{Increased Network Branching Heuristic.} 
\label{sec:INB} 

The principle of the increased network branching heuristic is derived from network reinforcement practice. In practice, network reinforcement works by connecting parts of the network directly to the transformer to reduce voltage drops and peak flows. For our purposes, this means to reconnect a subgraph $\widetilde{\Gamma}_j$ to a vertex of lower depth, such as the source or a vertex in the vicinity of the source. This reduces the depth of the network and increases the branching. By disconnecting $\widetilde{\Gamma}_j$ from a subgraph, material cost in this subgraph can be saved because the flows and voltage drops are getting smaller. These cost savings need to be compared to the additional cost for reconnecting $\widetilde{\Gamma}_j$ to a different vertex of the network. The pseudocode is provided in \Cref{alg:INB}. In \cref{line:INB_peca,line:INB_cost}, the capacities are determined and the cost of the initial layout is calculated. In \cref{line:save_layout}, the currently cheapest network layout is stored in $X^*$. The heuristic now loops over all vertices $j$ of a certain depth $d(j)$ up to a pre-defined maximum depth $d_{\mathrm{max}}$. In \cref{line:disconnect}, the heuristic disconnects the vertex $j$, resulting in two subgraphs: $\widetilde{\Gamma}_0$ and $\widetilde{\Gamma}_j$. In \crefrange{line:INB_loop2}{line:discard_inb}, the heuristic loops over all vertices $i \in \widetilde{\Gamma}_0$ with depth $d(i) < d$ and all vertices in $\widetilde{\Gamma}_j$ in order to evaluate potential cost reductions. In \cref{line:return_inb}, the heuristic returns the cheapest layout.


	\begin{algorithm}
	\SingleSpacedXI
		\footnotesize 
		\caption{Increased network branching heuristic} 																																					\label{alg:INB}
		\textbf{Input:} Network layout $\{x_{ij}\}$
		\begin{algorithmic}[1]
			\State $\{ a_{ij} \} \gets \textsc{CapacityOptimizationMethod}(\{x_{ij}\}) $																				\label{line:INB_peca}
			\State $c^* \gets \textsc{Cost}\left(\{x_{ij}\}, \{a_{ij}\}\right)$		\Comment{Determine initial cost}									\label{line:INB_cost}
			\State $X^* \gets \{x_{ij}\}$																																													\label{line:save_layout}
			\For{each vertex $j$ with $d(j) \in \{ 2, \ldots, d_{\mathrm{max}} \}$}																								\label{line:INB_loop}
				\State $x_{ij} \gets 0$   \Comment{Delete edge}    																											\label{line:disconnect}
        \State Compute subgraph $\widetilde{\Gamma}_j$ without source 
					\For{each vertex $p \not \in \widetilde{\Gamma}_j$ with $d(p) < d$}																									\label{line:INB_loop2}
						\For {vertex $q \in \widetilde{\Gamma}_j$}																																		\label{line:INB_recon}
							\State $x_{pq} \gets 1$			\Comment{Reconnect $\widetilde{\Gamma}_j$}
							\For{each edge $(m,n) \in \widetilde{\Gamma}_j$ in path from $j$ to $q$}
								\State $x_{mn} \gets 0, x_{nm} \gets 1$		\Comment{Check direction of reconnected edges}			\label{line:INB_direction}
							\EndFor
							\State $\{ a_{ij} \} \gets \textsc{CapacityOptimizationMethod}(\{x_{ij}\})$																\label{line:peca_inb}
							\State $c \gets \textsc{Cost}\left(\{x_{ij}\}, \{a_{ij}\}\right)$	\Comment{Determine new cost}									\label{line:cost_new_inb}
							\If{$c \leq c^*$} $X^* \gets \{x_{ij}\}$ and $c^* \gets c$ 			\Comment{Restore old layout if too expensive}	\label{line:opt_inb}
							\Else $\, \{x_{ij}\} \gets X^*$ 																																							\label{line:discard_inb}
							\EndIf		
						\EndFor
					\EndFor
			\EndFor
			\State \textbf{return} $X^*$																																													\label{line:return_inb}	
		\end{algorithmic}
	\end{algorithm}

The runtime of this heuristic depends on the network layout, in particular on the branching in the vicinity of the source. It further scales linearly with the runtime of the capacity optimization method in \cref{line:peca_inb}.

\subsubsection{Results} \label{sec:exp_setup}

We compare the performance of the following solution approaches:
\begin{enumerate}
\item The three heuristic solution approaches (Tabu Search, VNS, and INB).\footnote{For all heuristic solutions, the capacity optimization is carried out using the PECA heuristic. In \Cref{sec:add_info_p2}, we present a sensitivity analysis where we compare the PECA heuristic against alternative approaches for capacity optimization. Here, we find that the solution quality is on par with that of an exact solver, yet has a substantially lower runtime.} 
\item The performance of the linearized \pns
(as described in \Cref{sec:lin}) that is solved using the \emph{Gurobi MIP solver}. 
\item \NEW{The performance of the linearized \pns with Undiscounted Flows
(as described in \Cref{sec:commodities_model}) that is solved using the \emph{Gurobi MIP solver}.} 
\item We perform a \emph{complete enumeration} of layouts and capacities. Note that, for both MIP solver and enumeration approaches, in case an optimal solution could not be found within the given time limit, we evaluated the best solution obtained. 
\item We compute the \emph{upper} and \emph{lower bound} for the exact solution using the simplified problem instances (as described in \cref{sec:perf_eval_complete_problem}).
\end{enumerate}

The parameter configuration for these experiments is identical to the one presented in the main paper in \Cref{tab:parameters}. As in the main paper, to resemble real-world conditions, the experiments are conducted on network instances of various sizes $ N \in \{20, 40 \ldots, 100\}$. For each $N$, we generate 20 instances as follows and later report the averaged solution quality. The $x$- and $y$-locations of the vertices $(s_x, s_y)$ are sampled from a discrete uniform distribution without replacement.
The peak demand per household is set to $D^{\mathrm{peak}}=0.01$. For the coincidence factor, we use the model proposed by \cite{Rusck.1956}, \ie  $\gamma(|\Gamma_j|) = \gamma_{\lim} + (1-\gamma_{\lim}) \, |\Gamma_j|^{-1/2}$.

\Cref{tab:solution_comp} shows the solution quality for the various solution approaches. Note that the Gurobi solver and the enumeration approaches were not able to terminate within the given time limit of four hours. Instead, in case of the enumeration, we report the best solution found within the time limit; in case of the Gurobi solver, no feasible solution could be found within the given time limit \NEW{for neither the \pns nor the \pns with Discounted Flows}. We also experimented with much larger runtimes with the same result. 
\Cref{tab:solution_comp_time} shows the runtimes of the various solution approaches. The interpretation of the results is given in the manuscript in \cref{sec:perf_eval_layout} and \cref{sec:perf_eval_complete_problem}.

\begin{table}[ht]
\caption{Network cost depending on the number of loads $N$ for various solution approaches. \label{tab:solution_comp}}
\footnotesize
\centering
{
    \begin{tabular}{llrrrrrr}
    \toprule
    \textit{\textbf{Solution approach}} &       & \multicolumn{1}{l}{\textit{\textbf{N} $=$}} & \textbf{20} & \textbf{40} & \textbf{60} & \textbf{80} & \textbf{100} \\
    \midrule
    \multirow{3}{*}{Heuristics} & INB   &       & 17.5  & 37.5  & 57.1  & 76.5  & 105.4 \\
\cmidrule{2-8}          & VNS   &       & 16.3  & 34.9  & 53.9  & 78.3  & 104.0 \\
\cmidrule{2-8}          & Tabu Search &       & 15.8  & 34.2  & 52.3  & 76.2  & 103.6 \\
    \midrule
    \multirow{3}{*}{Exact approaches} & \textit{\pns \& Gurobi} &       & --$^{\dagger}$   & --$^{\dagger}$   & --$^{\dagger}$   & --$^{\dagger}$   & --$^{\dagger}$ \\
\cmidrule{2-8}          & \NEW{\textit{\pns with Disc. Flows} \& Gurobi} &       & \NEW{--$^{\dagger}$}  & \NEW{--$^{\dagger}$}  & \NEW{--$^{\dagger}$}  & \NEW{--$^{\dagger}$}  & \NEW{--$^{\dagger}$}  \\
\cmidrule{2-8}          & \textit{Complete enumeration} &       & 36.0$^{\ddagger}$ & 154.4$^{\ddagger}$ & 429.2$^{\ddagger}$ & 745.2$^{\ddagger}$ & 1057.8$^{\ddagger}$ \\
    \midrule
    \multirow{2}{*}{Bounds} & \textit{Lower bound} &       & 15.4  & 32.2  & 49.9  & 71.2  & 96.8 \\
\cmidrule{2-8}          & \textit{Upper bound} &       & 17.0  & 41.8  & 70.0  & 130.1 & 198.5 \\
    \bottomrule
    \multicolumn{8}{l}{\footnotesize $\dagger$ Calculation timed-out with no result.}\\
    \multicolumn{8}{l}{\footnotesize $\ddagger$ Best solution within time limit is shown as the calculation did not terminate.}\\
    \end{tabular}}
{}
\end{table}

\begin{table}[ht]
\caption{Average runtime in seconds depending on the number of loads $N$ for various solution approaches.}\label{tab:solution_comp_time}
\footnotesize
\centering
    \begin{tabular}{llrrrrrr}
    \toprule
    \textit{\textbf{Solution approach}} &       & \multicolumn{1}{l}{\textit{\textbf{N} $=$}} & \textbf{20} & \textbf{40} & \textbf{60} & \textbf{80} & \textbf{100} \\
    \midrule
    \multirow{3}{*}{Heuristics} & INB   &       & 2.9   & 208.9 & 2178.7 & 3319.2 & 39172.7 \\
\cmidrule{2-8}          & VNS   &       & 11.3  & 394.6 & 1490.5 & 4604.4 & 9320.4 \\
\cmidrule{2-8}          & Tabu Search &       & 13.1  & 232.9 & 835.4 & 2260.1 & 4654.5 \\
    \bottomrule
    \end{tabular}
{}
\end{table}

\NEW{Finally, we also investigated the use of decomposition techniques for solving the \pns. Specifically, we followed the Benders decomposition approach as recommended for MINLPs with integer and continuous variables \citep{conejo2006decomposition}. We refer to \citet{khodr2010optimal}, where this Benders decomposition is applied for a similar distribution network reconfiguration problem. However, the decomposition procedure does not result in simplifications for the \pns: Unlike in \citet{khodr2010optimal}, the \pns is composed of a very large number ($|V \times V \times K|$) of integer variables, which all remain in the master problem of the Benders decomposition. Therefore, the decomposition does not reduce problem complexity.}

\subsection{Capacity Optimization} 
\label{sec:add_info_p2}

In this section, we conduct a sensitivity analysis of the capacity optimization methods for cost and runtime. These results confirm that the PECA heuristic yields consistent results, independent from the network layout and demand. We repeat that the improvement heuristics trigger a capacity optimization for each candidate layout and, owing to this, the capacity optimization is responsible for a considerable part of the runtime of an improvement heuristic. 

We evaluate the PECA heuristic against an exact solution using the Gurobi Optimizer~7.5.2 and two greedy heuristics. In this evaluation, we use three different methods to generate network layouts: (1)~the MST, (2)~the greedy network construction, and (3)~a method whereby all edges are generated completely at random (random layout generation). The latter method generates a random Prüfer sequence of length $N - 2$, from which the network layout is created. For all these methods, the capacity optimization methods are evaluated in a low demand case and a high demand case. In total, this leads to six different test settings. In sum, these support our choice of the PECA heuristic, since, independent of how the network layout is constructed, it finds capacities that are close to the optimal solution yet in considerably less time.

Below, we first present the two greedy heuristics in more detail. Second and third, we present the results for the low and high demand experiments.

\subsubsection{Greedy Capacity Reinforcement.} 
\label{sec:greedy_reinforcement}

The greedy capacity reinforcement heuristic optimizes the capacities by steadily increasing them until the constraint for the voltage drops is fulfilled. It starts with minimal edge capacities, successively identifies the edge $(i,j)$ with the highest voltage drop (this corresponds to the weak spot of the network), and then increases the capacity of this edge. This heuristic resembles common industry practices in electricity network expansion. For instance, it is used by the partner company that provided us with real-world data. 

The heuristic works in five steps. In step~1, all capacities $a_{ij}$ are initialized to the minimum values that fulfill the line sizing constraint. At this point, it is not guaranteed that the constraint for the voltage drops in \Cref{eq:new_voltage} is fulfilled. 
In step~2, the heuristic calculates the voltage drops $\Delta_{U_{ij}}$ for each edge of the network. In step~3, the heuristic determines the set of paths $P' \subseteq P$ where the constraint for the voltage drops is violated. If all paths fulfill the constraints, the heuristic terminates and returns $\{a_{ij}\}$. In step~4, the heuristic considers all paths $p \in P'$ and reinforces the edge $(i,j) \in p$ with the highest voltage drop $\Delta_{U_{ij}}$ by increasing its capacity $a_{ij}$ to the next larger capacity, \ie from $a_{ij}^k$ to $a_{ij}^{k+1}$. In step~5, the voltage drop $\Delta_{U_{ij}}$ for this edge is recalculated, since increasing the capacity reduces the voltage drop. With these updated capacities, the heuristic returns to step~3. The runtime of this algorithm depends on the network layout. For example, in the case of a star network, the runtime is $O(N)$. For other layouts, runtimes are higher because both the number of paths and the depth of the paths increase with the network size $N$ \citepappendix[\cf][]{Steele.1987}.


\subsubsection{Greedy Capacity Reduction.} 
\label{sec:greedy_reduction}

The greedy capacity reduction heuristic is the counterpart to the greedy capacity reinforcement heuristic and proceeds in the opposite direction. It identifies the edge $(i,j)$ with the lowest voltage drop and reduces its capacity in order to save material cost. It works in five steps. In step~1, all capacities $a_{ij}$ are initialized to the maximum value possible and the heuristic creates a list $L$ containing all edges. In step~2, the heuristic calculates the voltage drops $\Delta_{U_{ij}}$ for each edge. In step~3, the heuristic identifies the edge $(i,j) \in L$ with the lowest voltage drop. If $L$ is empty, the heuristic terminates and returns $\{a_{ij}\}$. In step~4, the heuristic decreases $a_{ij}$ by one decrement from the value $a_{ij}^k$ to $a_{ij}^{k-1}$. In step~5, the heuristic evaluates whether this reduction violates the constraints related to line sizing and voltage drops. If they are violated, the capacity of the edge $(i,j)$ is reset to $a_{ij}^k$ and the edge is removed from $L$. If the constraints are still fulfilled, $\Delta_{U_{ij}}$ for this edge is recalculated and the heuristic returns to step~3.

The runtime of this heuristic scales similarly to the greedy capacity reinforcement heuristic. In practical applications, however, it entails a disadvantage with regard to runtime: close to the leaves, networks typically consist of many edges with low capacity. Thus, we expect its runtime to be slower than the runtime of the greedy capacity reinforcement heuristic as more iterations are required. 


\subsubsection{Results for Low Demand.}

We show the average cost per network for various instance sizes $N$ in \crefrange{tab:capacity_low_cost_MST}{tab:capacity_low_cost_random}. The PECA heuristic is largely on par with the exact solver (\ie Gurobi MIP solver), even for large networks. As expected, the PECA heuristic outperforms the other heuristics---greedy capacity reinforcement and greedy capacity reduction---due to its theoretical properties. 


In terms of runtime, the PECA heuristic outperforms the exact solver considerably, as dispayed in \crefrange{tab:capacity_low_runtime_MST}{tab:capacity_low_runtime_random}. Furthermore, the PECA heuristic has a slightly slower runtime than the greedy capacity reinforcement heuristic (but better cost performance as shown above). We observe that the greedy capacity reinforcement heuristic is computationally more efficient than the greedy capacity reduction heuristic. This matches our earlier expectations as solutions are likely to entail many edges with low capacity edges close to the leaves, which are more easily identified by the greedy capacity reinforcement heuristic compared to the greedy capacity reduction heuristic.


\begin{table}[ht!]
\caption{Cost sensitivity of capacity optimization to network layouts generated with MST.}
\label{tab:capacity_low_cost_MST}
{\resizebox{\columnwidth}{!}{ 
\renewcommand{\arraystretch}{1.0}
    \begin{tabular}{llrrrrrrrrrr}
    \toprule
    \textbf{Capacity Optimization} & \textit{\textbf{N}} & \textbf{5} & \textbf{10} & \textbf{15} & \textbf{20} & \textbf{25} & \textbf{30} & \textbf{35} & \textbf{40} & \textbf{45} & \textbf{50} \\
    \midrule
    Greedy capacity reinforcement &       & 7.43 (0.23) & 12.51 (0.16) & 16.46 (0.14) & 19.55 (0.11) & 22.25 (0.12) & 24.59 (0.11) & 27.77 (0.13) & 30.16 (0.11) & 31.62 (0.12) & 33.72 (0.11) \\
    \midrule
    Greedy capacity reduction &       & 7.43 (0.23) & 13.04 (0.19) & 17.65 (0.17) & 20.89 (0.13) & 23.66 (0.13) & 25.86 (0.12) & 28.81 (0.13) & 31.07 (0.10) & 32.46 (0.10) & 34.30 (0.09) \\
    \midrule
    PECA  &       & 7.41 (0.23) & 12.39 (0.15) & 16.12 (0.13) & 18.94 (0.10) & 21.44 (0.11) & 23.61 (0.10) & 26.45 (0.12) & 28.60 (0.11) & 30.04 (0.11) & 31.90 (0.10) \\
    \midrule
    Exact solver (Gurobi MIP) &       & 7.40 (0.22) & 12.37 (0.15) & 16.05 (0.13) & 18.85 (0.10) & 21.34 (0.11) & 23.48 (0.10) & 26.30 (0.12) & 28.41 (0.11) & 29.84 (0.11) & 31.63 (0.10) \\
    \bottomrule
    \multicolumn{12}{l}{Comparison of network cost using different capacity optimization methods for networks with various number of vertices $N$.}\\
    \multicolumn{12}{l}{Network layouts have been generated using the MST algorithm on randomly generated locations for each $N$.}\\
    \multicolumn{12}{l}{Figures shown are averages over 100 networks. The calculations were performed on an Intel Core i7-7600 CPU at 2.8 GHz and 16GB of RAM.}\\
    \end{tabular}}
}
{}
\end{table}

\begin{table}[ht!]
\caption{Cost sensitivity of capacity optimization to network layouts generated completely at random.}
\label{tab:capacity_low_cost_random}
{\resizebox{\columnwidth}{!}{ 
\renewcommand{\arraystretch}{1.0}
    \begin{tabular}{llrrrrrrrrrr}
    \toprule
    \textbf{Capacity Optimization} & \textit{\textbf{N}} & \textbf{5} & \textbf{10} & \textbf{15} & \textbf{20} & \textbf{25} & \textbf{30} & \textbf{35} & \textbf{40} & \textbf{45} & \textbf{50} \\
    \midrule
    Greedy capacity reinforcement &       & 12.00 (0.29) & 28.04 (0.21) & 44.92 (0.20) & 64.29 (0.16) & 84.80 (0.15) & 103.83 (0.15) & 122.39 (0.16) & 146.31 (0.15) & 170.41 (0.16) & 184.69 (0.15) \\
    \midrule
    Greedy capacity reduction &       & 12.05 (0.30) & 28.40 (0.21) & 45.48 (0.20) & 63.68 (0.15) & 83.08 (0.14) & 100.10 (0.13) & 117.28 (0.14) & 137.67 (0.12) & 156.74 (0.11) & 172.49 (0.11) \\
    \midrule
    PECA  &       & 11.99 (0.29) & 27.60 (0.21) & 43.44 (0.20) & 61.56 (0.15) & 79.93 (0.13) & 96.97 (0.13) & 113.72 (0.14) & 133.51 (0.12) & 151.31 (0.11) & 166.24 (0.11) \\
    \midrule
    Exact solver (Gurobi MIP) &       & 11.97 (0.29) & 27.51 (0.21) & 43.20 (0.19) & 61.13 (0.14) & 79.00 (0.13) & 95.62 (0.13) & 111.84 (0.13) & 131.06 (0.12) & 148.93 (0.11) & 163.62 (0.11) \\
    \bottomrule
    \multicolumn{12}{l}{Comparison of network cost using different capacity optimization methods for networks with various number of vertices $N$.}\\
    \multicolumn{12}{l}{Network layouts have been generated using a randomly generated layout with randomly generated locations for each $N$.}\\
    \multicolumn{12}{l}{Figures shown are averages over 100 networks. The calculations were performed on an Intel Core i7-7600 CPU at 2.8 GHz and 16GB of RAM.}\\
    \end{tabular}}
}
{}
\end{table}


\begin{table}[htb!]
\caption{Runtime sensitivity of capacity optimization to network layouts generated with MST.}
\label{tab:capacity_low_runtime_MST}
\centering
\footnotesize
    \begin{tabular}{llrrrrrrrrrr}
    \toprule
    \textbf{Capacity Optimization} & \textit{\textbf{N}} & \textbf{5} & \textbf{10} & \textbf{15} & \textbf{20} & \textbf{25} & \textbf{30} & \textbf{35} & \textbf{40} & \textbf{45} & \textbf{50} \\
    \midrule
    Greedy capacity reinforcement &       & 0.00  & 0.00  & 0.00  & 0.00  & 0.01  & 0.01  & 0.01  & 0.02  & 0.03  & 0.04 \\
    \midrule
    Greedy capacity reduction &       & 0.00  & 0.01  & 0.02  & 0.05  & 0.08  & 0.12  & 0.18  & 0.28  & 0.37  & 0.47 \\
    \midrule
    PECA  &       & 0.00  & 0.00  & 0.00  & 0.01  & 0.02  & 0.04  & 0.07  & 0.13  & 0.18  & 0.22 \\
    \midrule
    Exact solver (Gurobi MIP) &       & 0.01  & 0.02  & 0.03  & 0.06  & 0.09  & 0.12  & 0.16  & 0.23  & 0.27  & 0.33 \\
    \bottomrule
    \multicolumn{12}{p{13.5cm}}{Comparison of the average runtime using different capacity optimization methods for networks with various number of vertices $N$.}\\
    \multicolumn{12}{p{13.5cm}}{Network layouts have been generated using the MST algorithm on randomly generated locations for each $N$.}\\
    \multicolumn{12}{p{13.5cm}}{Figures shown are averages over 100 networks. The calculations were performed on an Intel Core i7-7600 CPU at 2.8 GHz and 16GB of RAM.}\\
    \end{tabular}
{}
\end{table}

\begin{table}[htb!]
\caption{Runtime sensitivity of capacity optimization to network layouts generated completely at random.}
\label{tab:capacity_low_runtime_random}
\centering
\footnotesize
{
    \begin{tabular}{llrrrrrrrrrr}
    \toprule
    \textbf{Capacity Optimization} & \textit{\textbf{N}} & \textbf{5} & \textbf{10} & \textbf{15} & \textbf{20} & \textbf{25} & \textbf{30} & \textbf{35} & \textbf{40} & \textbf{45} & \textbf{50} \\
    \midrule
    Greedy capacity reinforcement &       & 0.00  & 0.00  & 0.00  & 0.00  & 0.01  & 0.01  & 0.01  & 0.02  & 0.04  & 0.05 \\
    \midrule
    Greedy capacity reduction &       & 0.00  & 0.01  & 0.02  & 0.05  & 0.08  & 0.12  & 0.18  & 0.24  & 0.33  & 0.48 \\
    \midrule
    PECA  &       & 0.00  & 0.00  & 0.00  & 0.01  & 0.02  & 0.03  & 0.05  & 0.07  & 0.11  & 0.17 \\
    \midrule
    Exact solver (Gurobi MIP) &       & 0.01  & 0.02  & 0.03  & 0.06  & 0.09  & 0.12  & 0.17  & 0.21  & 0.26  & 0.37 \\
    \bottomrule
    \multicolumn{12}{p{13.5cm}}{Comparison of the average runtime using different capacity optimization methods for networks with various number of vertices $N$.}\\
    \multicolumn{12}{p{13.5cm}}{Network layouts have been generated using a randomly generated layout with randomly generated locations for each $N$.}\\
    \multicolumn{12}{p{13.5cm}}{Figures shown are averages over 100 networks. The calculations were performed on an Intel Core i7-7600 CPU at 2.8 GHz and 16GB of RAM.}\\
    \end{tabular}}
{}
\end{table}

\subsubsection{Results for High Demand.}

In \crefrange{tab:capacity_high_cost_MST}{tab:capacity_high_runtime_random}  we show average cost and average runtimes for the high demand case. For each $N$, we average over 100 networks. 

\begin{table}[ht]
\caption{Cost sensitivity of capacity optimization to network layouts generated with MST.}
\label{tab:capacity_high_cost_MST}
{\resizebox{\columnwidth}{!}{ 
\renewcommand{\arraystretch}{1.0}
    \begin{tabular}{llrrrrrrrrrr}
    \toprule
    \textbf{Capacity Optimization} & \textit{\textbf{N}} & \textbf{5} & \textbf{10} & \textbf{15} & \textbf{20} & \textbf{25} & \textbf{30} & \textbf{35} & \textbf{40} & \textbf{45} & \textbf{50} \\
    \midrule
    Greedy capacity reinforcement &       & 8.24 (0.24) & 13.77 (0.18) & 18.61 (0.15) & 22.52 (0.15) & 25.83 (0.13) & 29.45 (0.12) & 31.73 (0.12) & 34.25 (0.11) & 37.04 (0.11) & 38.90 (0.12) \\
    \midrule
    Greedy capacity reduction &       & 8.52 (0.27) & 14.65 (0.19) & 19.54 (0.15) & 23.01 (0.13) & 26.02 (0.11) & 28.94 (0.09) & 30.93 (0.09) & 33.28 (0.08) & 35.53 (0.08) & 37.13 (0.08) \\
    \midrule
    PECA  &       & 8.14 (0.24) & 13.25 (0.17) & 17.61 (0.15) & 21.13 (0.15) & 23.97 (0.12) & 26.87 (0.10) & 28.80 (0.10) & 30.93 (0.08) & 32.96 (0.08) & 34.40 (0.09) \\
    \midrule
    Exact solver (Gurobi MIP) &       & 8.10 (0.24) & 13.18 (0.17) & 17.51 (0.15) & 20.99 (0.15) & 23.78 (0.12) & 26.67 (0.10) & 28.53 (0.10) & 30.61 (0.09) & 32.63 (0.08) & 34.15 (0.09) \\
    \bottomrule
    \multicolumn{12}{l}{Comparison of network cost using different capacity optimization methods for networks with various number of vertices $N$.}\\
    \multicolumn{12}{l}{Network layouts have been generated using the MST algorithm on randomly generated locations for each $N$.}\\
    \multicolumn{12}{l}{Figures shown are averages over 100 networks. The calculations were performed on an Intel Core i7-7600 CPU at 2.8 GHz and 16GB of RAM.}\\
    \end{tabular}}
}
{}
\end{table}

\begin{table}[ht]
\caption{Cost sensitivity of capacity optimization to network layouts generated completely at random.}
\label{tab:capacity_high_cost_random}
{\resizebox{\columnwidth}{!}{ 
\renewcommand{\arraystretch}{1.0}
    \begin{tabular}{llrrrrrrrrrr}
    \toprule
    \textbf{Capacity Optimization} & \textit{\textbf{N}} & \textbf{5} & \textbf{10} & \textbf{15} & \textbf{20} & \textbf{25} & \textbf{30} & \textbf{35} & \textbf{40} & \textbf{45} & \textbf{50} \\
    \midrule
    Greedy capacity reinforcement &       & 13.63 (0.33) & 32.14 (0.22) & 53.20 (0.20) & 75.18 (0.19) & 95.91 (0.17) & 120.03 (0.15) & 143.08 (0.14) & 168.74 (0.15) & 190.08 (0.12) & 212.02 (0.13) \\
    \midrule
    Greedy capacity reduction &       & 13.92 (0.35) & 32.41 (0.22) & 51.54 (0.18) & 71.12 (0.16) & 89.15 (0.14) & 109.21 (0.12) & 128.65 (0.11) & 147.85 (0.12) & 168.24 (0.10) & 185.06 (0.11) \\
    \midrule
    PECA  &       & 13.42 (0.33) & 30.80 (0.21) & 49.39 (0.18) & 68.12 (0.16) & 85.36 (0.15) & 105.08 (0.13) & 123.79 (0.12) & 141.65 (0.12) & 162.13 (0.10) & 178.30 (0.11) \\
    \midrule
    Exact solver (Gurobi MIP) &       & 13.40 (0.33) & 30.60 (0.21) & 49.00 (0.18) & 67.32 (0.16) & 84.28 (0.15) & 103.62 (0.13) & 121.94 (0.12) & 140.05 (0.12) & 159.30 (0.10) & 175.37 (0.11) \\
    \bottomrule
    \multicolumn{12}{l}{Comparison of network cost using different capacity optimization methods for networks with various number of vertices $N$.}\\
    \multicolumn{12}{l}{Network layouts have been generated using a randomly generated layout with randomly generated locations for each $N$.}\\
    \multicolumn{12}{l}{Figures shown are averages over 100 networks. The calculations were performed on an Intel Core i7-7600 CPU at 2.8 GHz and 16GB of RAM.}\\
    \end{tabular}}
}
{}
\end{table}


\begin{table}[ht]
\caption{Runtime sensitivity of capacity optimization to network layouts generated with MST.}
\label{tab:capacity_high_runtime_MST}
\centering
\footnotesize
{
    \begin{tabular}{llrrrrrrrrrr}
    \toprule
    \textbf{Capacity Optimization} & \textit{\textbf{N}} & \textbf{5} & \textbf{10} & \textbf{15} & \textbf{20} & \textbf{25} & \textbf{30} & \textbf{35} & \textbf{40} & \textbf{45} & \textbf{50} \\
    \midrule
    Greedy capacity reinforcement &       & 0.00  & 0.00  & 0.00  & 0.01  & 0.01  & 0.02  & 0.02  & 0.03  & 0.05  & 0.06 \\
    \midrule
    Greedy capacity reduction &       & 0.00  & 0.01  & 0.02  & 0.04  & 0.07  & 0.09  & 0.15  & 0.20  & 0.26  & 0.34 \\
    \midrule
    PECA  &       & 0.00  & 0.00  & 0.01  & 0.02  & 0.04  & 0.05  & 0.08  & 0.10  & 0.15  & 0.19 \\
    \midrule
    Exact solver (Gurobi MIP) &       & 0.01  & 0.02  & 0.04  & 0.06  & 0.10  & 0.13  & 0.17  & 0.22  & 0.27  & 0.33 \\
    \bottomrule
    \multicolumn{12}{p{13.5cm}}{Comparison of the average runtime using different capacity optimization methods for networks with various number of vertices $N$.}\\
    \multicolumn{12}{p{13.5cm}}{Network layouts have been generated using the MST algorithm on randomly generated locations for each $N$.}\\
    \multicolumn{12}{p{13.5cm}}{Figures shown are averages over 100 networks. The calculations were performed on an Intel Core i7-7600 CPU at 2.8 GHz and 16GB of RAM.}\\
    \end{tabular}}
{}
\end{table}

\begin{table}[ht]
\caption{Runtime sensitivity of capacity optimization to network layouts generated completely at random.}
\label{tab:capacity_high_runtime_random}
\centering
\footnotesize
{
    \begin{tabular}{llrrrrrrrrrr}
    \toprule
    \textbf{Capacity Optimization} & \textit{\textbf{N}} & \textbf{5} & \textbf{10} & \textbf{15} & \textbf{20} & \textbf{25} & \textbf{30} & \textbf{35} & \textbf{40} & \textbf{45} & \textbf{50} \\
    \midrule
    Greedy capacity reinforcement &       & 0.00  & 0.00  & 0.00  & 0.01  & 0.01  & 0.02  & 0.03  & 0.04  & 0.05  & 0.06 \\
    \midrule
    Greedy capacity reduction &       & 0.00  & 0.01  & 0.02  & 0.04  & 0.07  & 0.10  & 0.14  & 0.19  & 0.25  & 0.31 \\
    \midrule
    PECA  &       & 0.00  & 0.00  & 0.01  & 0.01  & 0.02  & 0.04  & 0.06  & 0.08  & 0.10  & 0.13 \\
    \midrule
    Exact solver (Gurobi MIP) &       & 0.01  & 0.02  & 0.04  & 0.06  & 0.09  & 0.12  & 0.17  & 0.20  & 0.26  & 0.31 \\
    \bottomrule
     \multicolumn{12}{p{13.5cm}}{Comparison of the average runtime using different capacity optimization methods for networks with various number of vertices $N$.}\\
    \multicolumn{12}{p{13.5cm}}{Network layouts have been generated using a randomly generated layout with randomly generated locations for each $N$. Figures shown are averages over 100 networks.}\\
    \multicolumn{12}{p{13.5cm}}{The calculations were performed on an Intel Core i7-7600 CPU at 2.8 GHz and 16GB of RAM.}\\
    \end{tabular}}
{}
\end{table}

\section{Additional Information on Computational Experiments}
\label{sec:add_info_comp}


The following parameters are used for the improvement heuristics: The increased and decreased branching heuristics use the parameter $d_{\mathrm{max}} = 4$. The VNS algorithm uses $d_{\text{max}} = 3$, $s_{\text{VNS}} = 5$, and $s_{\text{max}} = N$ for the local search. The Tabu Search algorithm uses $s_{\text{tabu}} = 10\,N$. For instances with $N \leq 20$, the length of the tabu list is set to $5$; for all other instances, it is set to $10$. 

The heuristics are implemented in Python~3.5. All computational experiments are conducted in parallel on 4 multi-core Intel~Xeon~E5-2630~v4~CPUs at 2.2\,GHz and 8\,GB of RAM. 16 computations are running in parallel at any given time. We checked that this process does not impair the individual runtimes. Note that, for the exact solution methods, we experimented with large runtime limits of several days; yet, due to the problem complexity, exact solutions are prevented.

\section{Additional Information on Real-World Case Study}
\label{sec:add_info_case}


Below, we report objective function and main constraints under the parameter configuration and units used for the real-world case study. 
The objective function reads
\begin{align}
 \sum\limits_{(i, j)^k \in E} x_{ij}^k \, [34.62 \, \tfrac{\mathrm{CHF}}{\mathrm{m}} \, l_{ij} + 0.1882 \, \tfrac{\mathrm{CHF}}{\mathrm{m \; mm^2}} \, l_{ij} \, a_{ij}^k] \, .
\end{align}
\noindent
The line sizing constraint for our case study reads 
\begin{align}
		\SI{0.4}{kV}  \sum \limits_{k \in \{1, \ldots, |A|\}} x_{ij}^k c_{ij}^{k} \geq  F_{ij} \, ,		\quad  \forall i, j \in \{0, \ldots, n - 1 \} \, . 
\end{align}
\noindent
The constraint for the voltage drops reads
\begin{align}
	\newsubeqblock
	\mysubeq & && \sum\limits_{k} x_{ij}^k \frac{a_{ij}^k}{l_{ij}} (U_i - U_j) = \tfrac{\sqrt{3}}{\SI{0.4}{kV}} \; 0.0181 \tfrac{\Omega \mathrm{mm}^2}{\mathrm{m}} F_{ij} \, , \quad  \forall i, j \in \{0, \ldots, N - 1 \}  \, ,     \label{eq:constr_voltage_con_1}
	\\
	\mysubeq & && U_i \geq \SI{0.4}{kV} - \SI{3}{\percent} \; \SI{0.4}{kV} \, , \quad \forall  i \in \{0, \ldots, N - 1 \} \, ,     \label{eq:constr_voltage_con_2}
		\\
	\mysubeq & && U_0 =  \SI{0.4}{kV}  \, .     \label{eq:constr_voltage_con_3}
\end{align}
\noindent

\SingleSpacedXI
\bibliographystyleappendix{informs2014} 
\bibliographyappendix{literature, grid_planning_gg,literature_revision}

\end{document}